\documentclass[journal=jpccck,manuscript=article,layout=twocolumn]{achemso}
\usepackage{chemformula}
\usepackage[T1]{fontenc}
\usepackage{amsmath}
\usepackage{amssymb}

\usepackage{epsfig,bm}

\author{Francesco Cordero}
\email{francesco.cordero@ism.cnr.it}
\author{Floriana Craciun}
\affiliation[ARTOV]
{Istituto di Struttura della Materia-CNR (ISM-CNR), Area della Ricerca di Roma - Tor Vergata, Via del Fosso del Cavaliere 100, I-00133 Roma, Italy}
\author{Francesco Trequattrini}
\affiliation[Fis]
{Dipartimento di Fisica, Universit\`{a} di Roma "La Sapienza", p.le A. Moro 2, I-00185 Roma, Italy}
\alsoaffiliation[ARTOV]
{Istituto di Struttura della Materia-CNR (ISM-CNR), Area della Ricerca di Roma - Tor Vergata, Via del Fosso del Cavaliere 100, I-00133 Roma, Italy}
\author{Amanda Generosi}
\author{Barbara Paci}
\affiliation[ARTOV]
{Istituto di Struttura della Materia-CNR (ISM-CNR), Area della Ricerca di Roma - Tor Vergata, Via del Fosso del Cavaliere 100, I-00133 Roma, Italy}
\author{Anna Maria Paoletti}
\author{Gloria Zanotti}
\affiliation[MLIB]
{Istituto di Struttura della Materia-CNR (ISM-CNR), Area della Ricerca di Roma 1, Via Salaria, Km 29.300, I-00015 Monterotondo Scalo, Roma, Italy}

\title{Influence of Temperature, Pressure and Humidity on the Stabilities and Transitions Kinetics of the Various Polymorphs of FAPbI$_3$ }

\begin{document}


\begin{tocentry}
\includegraphics[width=8.5 cm]{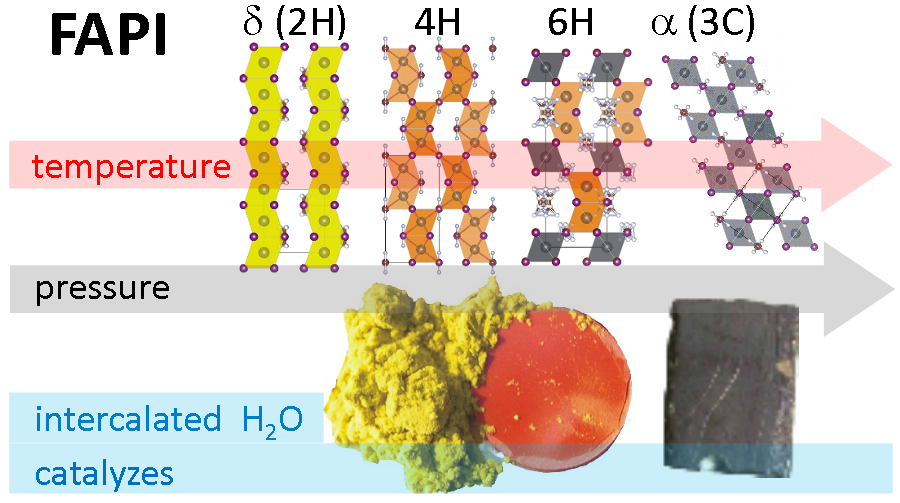}
\end{tocentry}

\begin{abstract}
The phase transitions between the various polymorphs of FAPbI$_{3}$ (FAPI,
FA = formamidinium CH(NH$_{2}$)$_{2}^+$) are studied by anelastic, dielectric
and X--ray diffraction measurements on samples pressed from $\delta -$FAPI
(2H phase) yellow powder. The samples become orange after application of as
little as 0.2~GPa, which has been explained in terms of partial
transformations to the other hexagonal polymorphs 4H and 6H. The phenomenon
is discussed in the light of what is known about the stability of the
various polymorphs of hybrid and inorganic perovskites ABX$_{3}$ with large
A cations and hence large tolerance factor $t$. Remarkably, FAPI at room and
higher temperature behaves like a perovskite with large $t$, while just
below room temperature it behaves like a perovskite with small $t$. The
kinetics of the transformations between the polymorphs is enhanced by small
amounts of intercalated water. It seems therefore worthy to try improving
the atomic diffusion and crystallization during synthesis, and hence the
final photovoltaic performance, through controlled small amounts of water that
should be thoroughly removed after a sufficiently homogeneous and smooth
microstructure is achieved.
\end{abstract}

\section{Introduction}

The hybrid metalorganic and the inorganic halide perovskites, ABX$_{3}$ with
A = organic cation or Cs$^+$, B\ = Pb$^{2+}$, Sn$^{2+}$, X = I$^-$, Br$^-$, Cl$^-$,
have been the object
of extensive investigations since it was demonstrated that they exhibit
excellent photovoltaic performance and may enable the creation of solar cells%
\cite{GHS14,CSB17} and optoelectronic devices\cite{SS16b} with low costs and
high performance. What prevents the commercialization of devices based on
these materials is their poor stability, especially in the presence of
humidity, light and heat, all factors that are especially crucial in
photovoltaic applications.\cite{CSB17,AMM19,BCL19}

The main directions for improving the phase stabilities and power conversion
efficiencies are compositional engineering or use of additives or surface
protecting layers, but the effect of pressure is also explored,\cite%
{LKY19,LLW20} since effects similar to the application of pressure may be
obtained by shrinking the lattice of the perovskite film through the choice
of the substrate.\cite{KLG16,GAD19} In some cases it has been reported that
the application of pressure and even of humidity during preparation have
positive roles in improving the stability or morphology and photovoltaic
efficiency of films of these perovskites.

The main effect of pressure in perovskites with rigid B--X bonds is
promoting tilting of the BX$_{6}$ octahedra. This occurs in order to relieve
the mismatch between the shrinking A-X framework and the rigid BX$_{6}$
octahedra, which is usually quantified by how much the Goldschmidt's
tolerance factor,\cite{Gol26,Meg52} defined in terms of the equilibrium
ionic radii as
\begin{equation}  \label{eq-t}
t=\frac{r_{\mathrm{A}}+r_{\mathrm{X}}}{\sqrt{2}\left( r_{\mathrm{B}}+r_{%
\mathrm{X}}\right) }~,
\end{equation}%
is less than 1. In fact, $t=$ $1$ for radii that perfectly fit the cubic
perovskite; when $t<$ $1$ the BX$_{6}$ octahedra are too large to fit the
A--X bonds and therefore, if they are also more rigid than the A--X bonds,
they tilt without shrinking. In the hybrid perovskites, however, the H bonds
between the A organic molecule and the X anions also play a role.\cite{LBL16}
Much less clear is the role of pressure in promoting the transitions between
the various polymorphs of the halide perovskites with different octahedral
coordinations. This is particularly true for the case of FAPbI$_{3}$ (FAPI),
for which the literature on the effects of pressure is contradictory.\cite%
{LKY19} The topic is of particular relevance, since the other polymorphs of
these perovskites are also the phases into which they generally transform
during degradation, being more stable in ambient conditions. These phases
have yellow or orange colors and definitely lower photovoltaic performance.
The various phases are often labeled as $\alpha $ for the black cubic
perovskite, which at lower temperature transforms into $\beta $ and $\gamma $
with tilted BX$_{6}$ octahedra and $\delta $ for the yellow non-perovskite,%
\cite{SMK13c,SK15} though recently it has been shown that a series of
intermediate polymorphs of FAPI exist.\cite{GZS17}
They are recognizable by diffraction and from the color, or more
quantitatively from the absorption bands in the UV-visible region.%
\cite{GZS17,Gra18,SMS19,DWG19}
In this context Ramsdell's notation\cite{Ram1947} is useful, according to
which the cubic perovskite phase is labeled 3C and the $\delta - $FAPI phase
is 2H, while the intermediate phases are 4H and 6H.
All first-principle
calculations indicate that the lowest energy and therefore stable polymorph
of FAPI is the $\delta $ phase, and indeed this is the phase obtained at
room temperature with chemical methods. Yet, dry grinding or ball milling
the starting FAI and PbI$_{2}$ powders may produce the perovskite phase,\cite%
{SMK13c,DLL18,HTJ19} and similarly for MAPbI$_{3}$\cite{MYM16} (MAPI, MA =
methylammonium (CH$_3$NH$_3$)$^+$). Although a relationship between the
application of high pressure and mechanosynthesis is not recognized, it can
be estimated that in ball milling the impacting balls generate pressures of
several GPa in the contact region.\cite{LRC16}

There are also reports of new phases, also with improved photovoltaic
properties and stabilities, obtained after the application of high pressure.
These phases generally have shorter bonds, higher orbital overlap and
reduced bandgap, resulting in enhanced photocurrents. They are proposed to
be accessible after pressure--induced amorphization and subsequent
recrystallization, as in the case of tetragonal MASnI$_{3}$, which becomes
cubic,\cite{LWS16} or the 2D perovskite (BA)$_{2}$(MA)$_{3}$Pb$_{4}$I$_{13}$
(BA$^+$~= butylammonium (CH$_3$(CH$_2$)$_3$NH$_3$)$^+$).\cite{LGK18} In FAPI a
similar improvement of the bond configuration accompanied by bandgap
decrease has been reported to persist after removal of a pressure of
2.1~GPa, rather surprisingly, even in the absence of any intermediate
pressure--induced phase transition or amorphization.\cite{LKG17} However,
there are contrasting results on the effect of pressure on $\alpha - $FAPI.%
\cite{PM17} It has also been found that as little as 0.1~GPa would promote
the $\alpha \rightarrow \delta $ transformation,\cite{JLJ18} while a single
crystal has been found to undergo at 0.49~GPa a transition to the tetragonal
$P4/mbm$ structure, due to octahedral tilting,\cite{SDW17} presumably the
same $\beta $ phase that is obtained upon cooling below 285~K. Finally,
pressures of 4-6~GPa have been found to reversibly turn the black and yellow
phases into red,\cite{WGG17} and piezochromism has been found also in
nanocrystalline FAPI\cite{ZCQ18} and in FAPbBr$_{3}$,\cite{WWZ16}
interpreted in terms of the dependence of the bandgap on the Pb--I--Pb bond
lengths and angles in the variously tilted perovskite phases.\cite{ZCQ18}

Also humidity affects FAPI and generally metal--organic perovskites in
different manners. Besides the well established role as catalyst of the
degradation of the perovskite $\alpha $ phases into non-perovskite $\delta $
and final decomposition, a relative humidity (RH) of 30\% has been found to
generally improve the quality of films compared to the dry process \cite%
{ZCL14} and particularly to improve the crystallization and enlarge the
grain size of MAPI and FAPI films,\cite{HTL17}. Also in the mechanosynthesis
of MAPI a RH of 40\% accelerates the reaction between MAI and PbI$_{2}$ .%
\cite{MYM16}

Here we present dielectric, anelastic and x--ray diffraction measurements on
samples compacted from $\delta -$FAPI powder. The dielectric and elastic
properties are particularly sensitive to the level of hydration and volume
fractions of $\alpha $ and $\delta $ phases, since dry $\alpha -$FAPI has
elastic modulus and dielectric constant at least twice as large as the $%
\delta $ and hydrated phases. These experiments shed light and allow some
rationalization to be made on the behavior of the several polytypes of FAPI
under pressure and humidity.

\section{Methods}

We performed dielectric, anelastic and X-ray diffraction (XRD) measurements
on samples compacted from the yellow $\delta-$FAPI powder.
The powder was obtained following a procedure already reported\cite{LSY16}
with slight modifications, precipitating with toluene the $\delta $--FAPI
phase from a solution of HC(NH$_{2}$I and PbI$_{2}$ in GBL. The
yellow powder was recovered by centrifugation and dried at room temperature
in vacuum overnight as described in the Supporting Information (SI). The
powder for one of the bars (FAPI \#7) was prepared in water following the
procedure reported in Ref. \citenum{JBN19} with slight modifications, as
detailed in the SI, obtaining pure $\delta -$FAPI according to the XRD
analysis.

The bars for the anelastic and discs for the dielectric measurements were obtained by
pressing the FAPI powder for few minutes at maximum pressures of 0.2 and
0.62~GPa respectively in dies with section $40\times 6$~mm$^{2}$ or 13~mm
diameter and their thicknesses were between 0.5 and 1~mm.

The complex Young's modulus $E = E' + i E''$  versus temperature was measured in a high
vacuum (HV) insert with base pressure $<10^{-6}$~mbar, with the introduction of
$<0.2$~mbar He below room temperature for thermal exchange or in controlled
H$_2$O pressure. The FAPI thin bars were suspended
on thermocouple wires and electrostatically excited on their flexural modes
as described in Ref. \citenum{CDC09}.
The real part of the complex Young's modulus is proportional to the square
of the fundamental resonance frequency $f$ of the first flexural mode as\cite{NB72}
\begin{equation*}
f=1.028\frac{h}{l^{2}}\sqrt{\frac{E}{\rho }}~,
\end{equation*}%
where $l$, $h$, $\rho $ are the sample's length, thickness and density.
The elastic energy loss $Q^{-1} = E''/E'$ was measured from the width of
the resonance peaks or the free decay.

The dielectric permittivity $%
\varepsilon $ versus temperature was measured in a small closed volume, a
modified Linkam cell, that does not allow HV tight conditions.
Previous experience taught us that the standard
purging procedure of few minutes with dry N$_{2}$ flux at 40~${{}^{\circ }}$%
C is totally ineffective in removing the water absorbed by FAPI.\cite{CCT19}
In addition, at low temperature some humidity may leak into the cell and
condense on the sample, giving rise to a broad peak of $\varepsilon $ during
heating back to room temperature. Therefore, in order to remove water as
efficiently as possible from the sample, purging was made at the highest
temperature possible under the given conditions at the beginning and at the
end of each temperature run.

The XRD spectra were acquired in air at various stages of the anelastic
and dielectric experiments and on freshly prepared discs, as described
in more detail in the SI.

\section{Results}

The samples
pressed from the yellow $\delta -$FAPI were orange. The discs, with smaller
surface and subjected to higher pressure, were of more uniform and intense
color, while the bars presented lighter zones corresponding to regions where
the initial thickness of the powder and therefore local pressure and final
density were smaller (see Figs. \ref{S1} and  \ref{S2}(a) in SI). We understand this
phenomenon as a pressure--induced transformation from the yellow 2H phase to
the orange and red 4H and 6H phases,\cite{GZS17,Gra18} more complete in the
case of the discs subjected to higher pressure. The presence of these phases
is evident from the color but does not give clear signatures in the XRD
spectra, as shown in the SI.

\begin{figure}[tbh]
\includegraphics[width=8.5 cm]{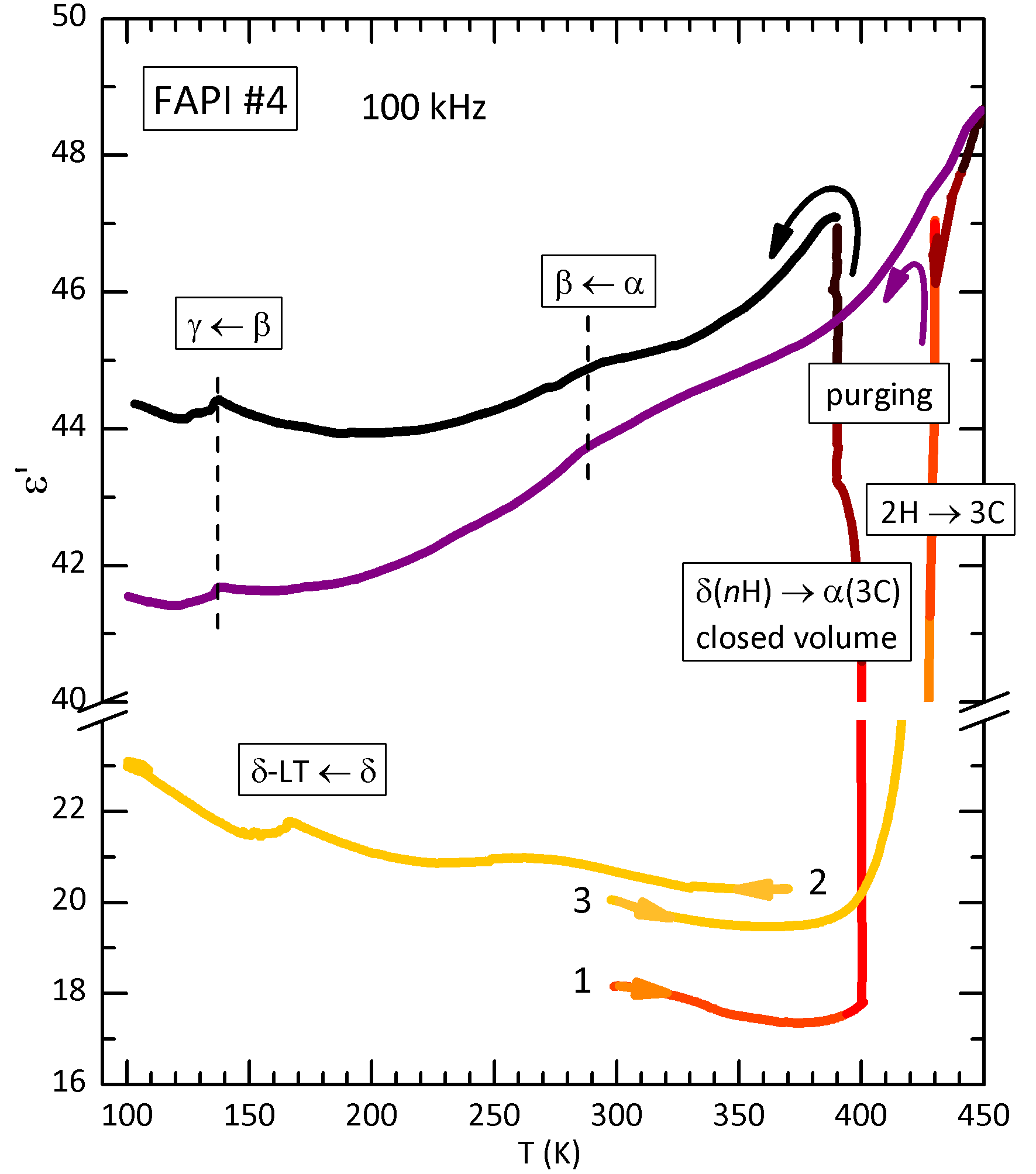}
\caption{{}Dielectric permittivity of the disc FAPI \#4 pressed from powder
of $\protect\delta -$FAPI measured at 100~kHz during several temperature
runs (only few are shown). The colors are indicative of the state of the
sample; the yellow color of the sample in curve 2 after three days at room
temperature in the measurement cell corresponds to the $\protect\delta$
phase but has not been checked.}
\label{fig-diel}
\end{figure}

Figure \ref{fig-diel} shows the dielectric permittivity of an
initially orange disc pressed from yellow $\delta-$FAPI.

\begin{figure}[tbh]
\includegraphics[width=8.5 cm]{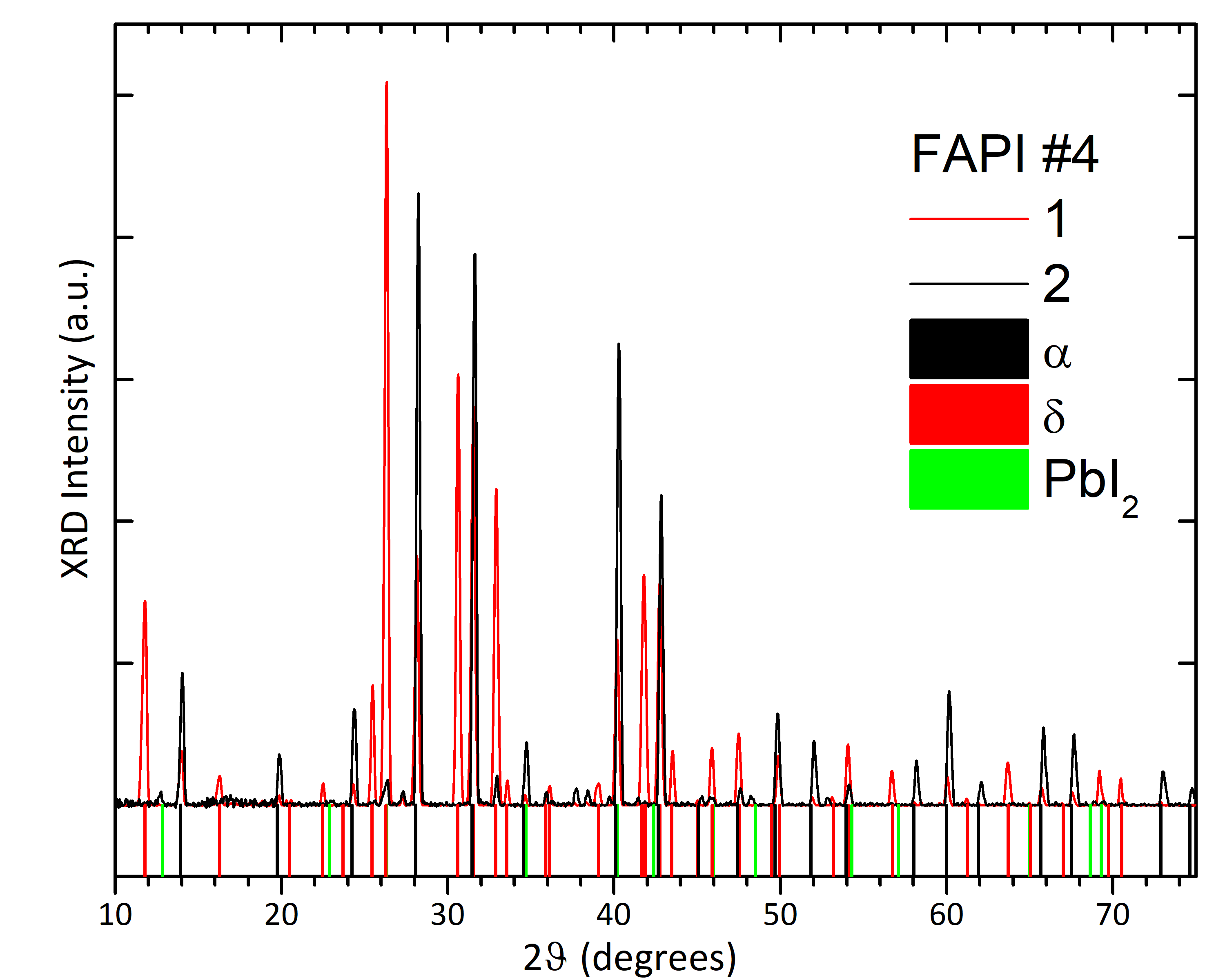}
\caption{{} XRD pattern collected on sample FAPI \#4 after the dielectric
measurements in Fig. \ref{fig-diel} (curve 1) and after further temperature
runs including longer annealing at 430--440~K.
Black, red and green lines are the expected reflections of the $\alpha $,
$\delta $ and PbI$_2$ phases respectively.}
\label{fig-XRD-FAPI4}
\end{figure}

We started the first heating (curve 1) with the intention of
stopping temperature immediately after
the dielectric permittivity started to rise, indicating the onset of the $n$H%
$\rightarrow $3C transition, where $n$H indicates some mixture of 2H, 4H and
6H phases. Such a transition, however, was so fast that we were not able to
prevent the formation of 3C phase, as indicated by the $\alpha \rightarrow
\beta $ and $\beta \rightarrow \gamma $ transitions measured during the
subsequent cooling.\cite{CCT19} The final purging at 380~K was apparently
insufficient to remove water from the sample, which after three days (curve
2) was back in the $\delta $ (presumably 2H)\ phase with low permittivity,
no trace of the $\alpha \rightarrow \beta $ and $\beta \rightarrow \gamma $
transitions, and instead the transition to the LT-$\delta $ phase known to
occur at 173~K upon heating (167~K in the cooling curve 2) after thermal
expansion\cite{KOW19} and neutron diffraction.\cite{CFP16} A new temperature
run (curve 3) was extended to 450~K, but this time the 2H$\rightarrow $3C
transition was slower than in curve 1, and, even though a 12 min purging was
done at 430~K, the subsequent cooling did not show signs of a more complete
transition to the $\alpha $ phase with respect to curve 1. On the contrary,
the permittivity was slightly lower and the transitions less marked. After
the series of dielectric measurements the sample had a shiny black surface,
with some dark violet reflections, but no signs of cracks, indicating that
the transitions between the $n$H and 3C phases occur in a nearly congruent
manner without excessive internal strains. The XRD spectra (pattern 1 of Fig.
\ref{fig-XRD-FAPI4}) contained only the peaks of the 3C and 2H phases; the
latter was evidently induced by humidity and its presence is in agreement
with the lower permittivity during the last dielectric run.

After further dielectric runs including a prolonged aging of 110~min
at 430--440~K (not shown) the XRD reflections of the $\alpha$ phase were
found more intense, but also some refections of PbI$_2$ started being
detected (pattern 2).

Even if taken within one day after a dielectric or anelastic experiment,
the XRD spectra do not capture the status of the sample at the end of that
experiment, as is clear from the near total transformation of these compacted
samples from the $\alpha $ to the $\delta $ phase in only three days in
presence of humidity (Fig. \ref{fig-diel}). Yet, we noted several times
an overemphasized appearance of the XRD peaks of the hexagonal
$\delta $ and PbI$_2$ phases over the $\alpha $ phase with respect to
the last anelastic or dielectric experiment and the sample appearance.
We speculate that this is due to a superior crystallinity of the hexagonal
phases over cubic $\alpha -$FAPI, as explained later on.

We pass to the description of the results of the anelastic spectra
obtained during several temperature runs in HV conditions ($10^{-6}-10^{-5}$%
~mbar or $\sim 10^{-2}-10^{-1}$~mbar He below room temperature).

\begin{figure}[tbh]
\includegraphics[width=8.5 cm]{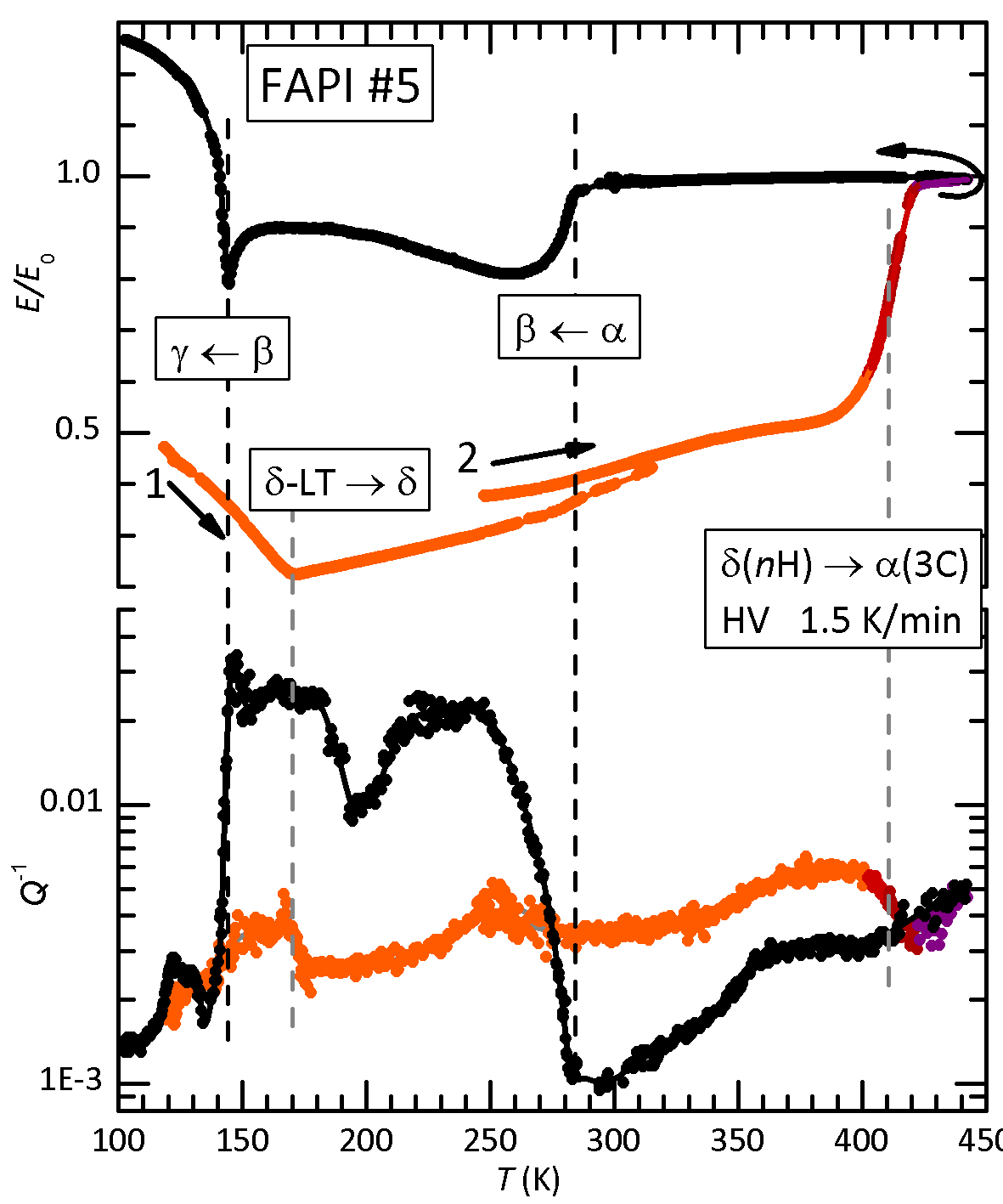}
\caption{{}Young's modulus and elastic energy loss coefficient $Q^{-1}$ of a
bar pressed from powder of $\protect\delta -$FAPI during several temperature
runs (only few are shown).}
\label{fig-FAPIn5}
\end{figure}

Figure \ref{fig-FAPIn5} shows few temperature runs on sample FAPI \#5, an
initially orange bar (fig. \ref{S2}(a)) obtained by pressing yellow powder of
freshly prepared $\delta -$FAPI. The initial state is therefore a mixture of
2H, 4H and possibly 6H and similar phases. The upper panel is the relative
change of the Young's modulus $E$, deduced from the resonance frequency.
The reference $E_{0}$ corresponds to $f=$ 2.0~kHz in the $\alpha $ phase. Its
absolute value can be estimated as $8.5$~GPa but with a large error, of the
order of 20\%, due to the non uniform thickness and density of the bar. The
lower panel is the elastic energy loss coefficient $Q^{-1}=E^{\prime \prime
}/E^{\prime }$. Below 380~K the major feature of the anelastic spectrum of $%
\delta -$FAPI is the phase transformation at $T_{\mathrm{LT}}=$ 170~K, also
visible in the dielectric spectrum (curve 2 of Fig. \ref{fig-diel}). We
label $\delta -$LT the phase below 170~K, which has been shown to cause a
lattice expansion\cite{KOW19} and the appearance of new Bragg peaks\cite%
{CFP16} and has a reduced number of possible orientations of the FA
molecules.\cite{CCF17c} The Young's modulus softens almost linearly at both
sides of the minimum at $T_{\mathrm{LT}}$, with higher slope on the low
temperature side, suggesting a proper or at least pseudoproper ferroelastic
transition,\cite{Wad82} where a strain component is the order parameter or
has its symmetry and the corresponding compliance has a maximum of the
Curie-Weiss type.\cite{CS98} Notice that the minimum in the softened elastic
constant is more marked than it appears in the polycrystalline Young's
modulus $E$, which contains the contribution of the other elastic constants
of the 2H phase not involved in the transition and of the other $n$H phases,
which presumably do not undergo the same transition. A similar elastic
anomaly has been observed in the metal--organic framework perovskite [(CH$%
_{2}$)$_{3}$NH$_{2}$]Mn(HCOO)$_{3}$ in correspondence with a phase
transition where the azetidinium molecules undergo orientational ordering,%
\cite{LZB13} which probably is similar to what occurs to the formamidinium
molecules in the $\delta -$LT phase of FAPI. In that case, in order to
maintain that the coupling between strain and order parameter is
linear-quadratic, the linear increase of the elastic constant above the
transition temperature has been interpreted as due to coupling between
strain and strong fluctuations of the order parameter, while the stiffening
below the transition was explained supposing that the transition is close to
tricritical.\cite{LZB13} Indeed, there is little hysteresis, $\leq 2$~K, in $%
T_{\mathrm{LT}}$ between heating and cooling, but regarding possible
fluctuations above $T_{\mathrm{LT}}$, they do not seem to appear in the
dielectric permittivity, which just has a sharp step below $T_{\mathrm{LT}}$.
In addition, there is a marked hysteresis in the value of the minimum of $E$,
since the material is softer during heating and partially recovers at
room temperature (not shown). This accounts for the fact that curve 2,
measured after a few cycles at low temperature, does not coincide with curve
1. Curve 2 was measured after the sample had been 6 days in the insert for
anelasticity measurements in HV or static He atmosphere and therefore the
sample was completely dehydrated (at 340~K the vacuum was $7\times 10^{-7}$%
~mbar). Therefore curve 2 represents the transition from the $\delta $
phase, actually a mixture of 2H, 4H and possibly other hexagonal phases, to
the 3C cubic $\alpha $ phase during heating at 1.5~K/min in totally
anhydrous condition. The transition occurs between 400 and 420~K. Heating
was stopped at 443~K since no further stiffening was observed, indicating
that the transformation was completed, and in order to avoid the sample
decomposition. In this condition the Young's modulus and losses were
perfectly stable and the cooling curves were identical to those previously
measured on FAPI,\cite{CCT19} with the $\alpha \rightarrow \beta $
transition accompanied by a negative step in the modulus and rise of
dissipation, and the $\beta \rightarrow \gamma $ transition accompanied by a
seemingly critical softening followed by marked stiffening well above the
value in the $\alpha $ phase.


\begin{figure}[tbh]
\includegraphics[width=8.5 cm]{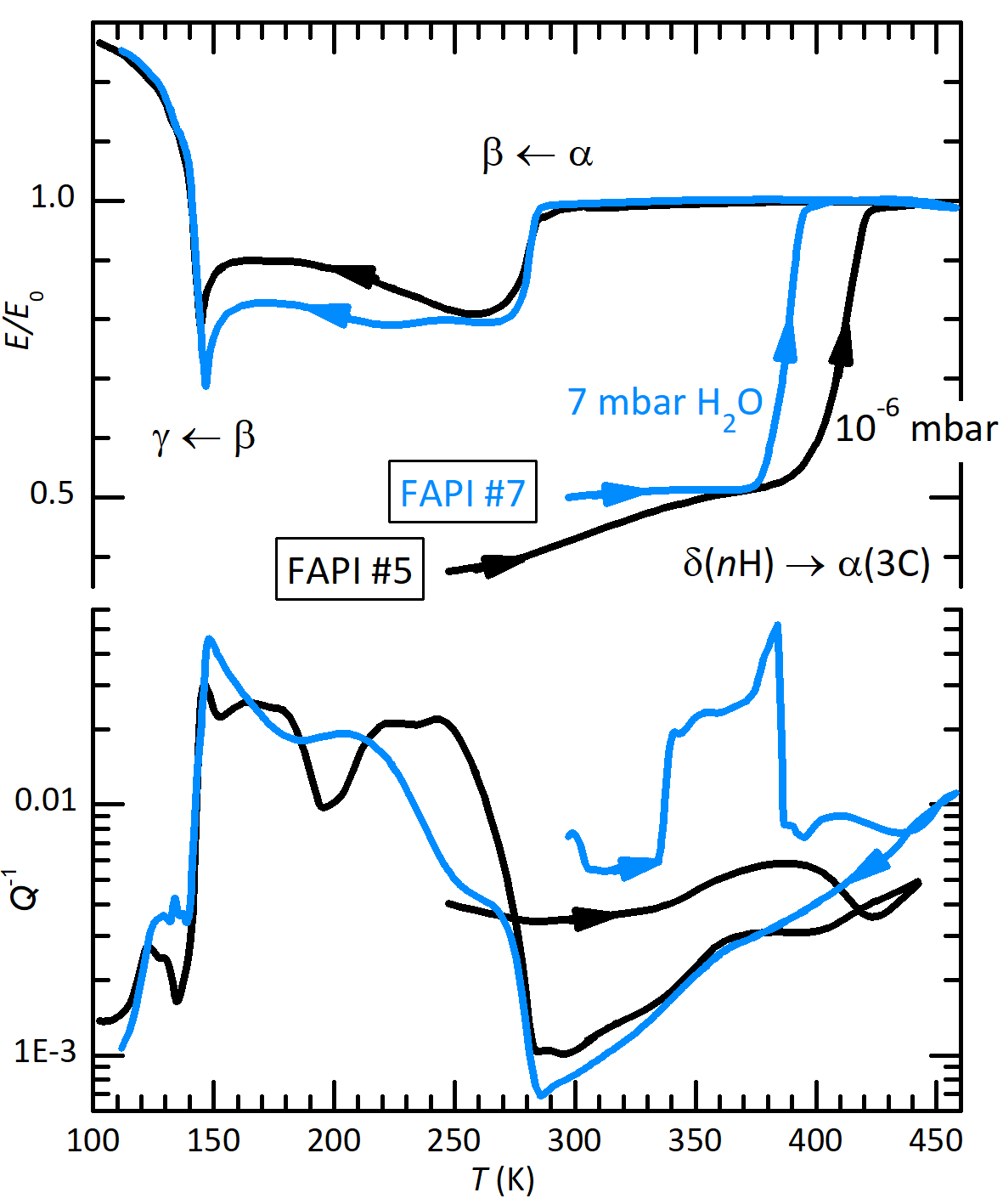}
\caption{{}Young's modulus and elastic energy loss of two bars $\protect%
\delta -$FAPI (mixture of $n$H phases) heated in completely anhydrous
condition (FAPI \#5, same as Fig. \protect\ref{fig-FAPIn5} ) and under
7~mbar H$_{2}$O (FAPI \#7). After the transformation to the 3C $\protect%
\alpha $ phase, the transitions to the $\protect\beta $ and $\protect\gamma $
structures are observed during cooling. In both cases the heating rate was
1.5~K/min.}
\label{fig-FAPI5-7}
\end{figure}

The dielectric experiments suggested that the transition to the cubic black
3C phase was faster for the orange $n$H ($n>2$) phases induced by pressure
than for the yellow 2H phase, but the above anelastic experiments on an
orange sample did not show the sudden transition to 3C at $\leq 400$~K. In
order to check whether also the content of water has a role in the kinetics
of these structural transitions, we prepared a new bar (FAPI \#7) from
yellow powder and heated it at 1.5~K/min in 7~mbar H$_{2}$O (up to 445~K and
then HV as usual; 7~mbar correspond to a relative humidity of 30\% at room
temperature). The comparison with the previous measurement on FAPI \#5,
completely dehydrated and heated in a vacuum of $10^{-6}$~mbar at the same
temperature rate, is presented in Fig. \ref{fig-FAPI5-7}. The rise of the
modulus indicating the transition to the $\alpha $ phase occurs in advance
of about 24~K in the presence of water and, judging from the amplitudes of
the elastic anomalies at the $\alpha \rightarrow \beta $ and $\beta
\rightarrow \gamma $ transitions, the transformation to the $\alpha $ phase
is more complete in the hydrated sample. There is also a sudden rise of
dissipation when heating through 334~K, which disappears at the $\delta
\rightarrow \alpha $ transition.

\section{Discussion}

\subsection{Relative volumes of the FAPI polymorphs}

\begin{figure}[tbh]
\includegraphics[width=8.5 cm]{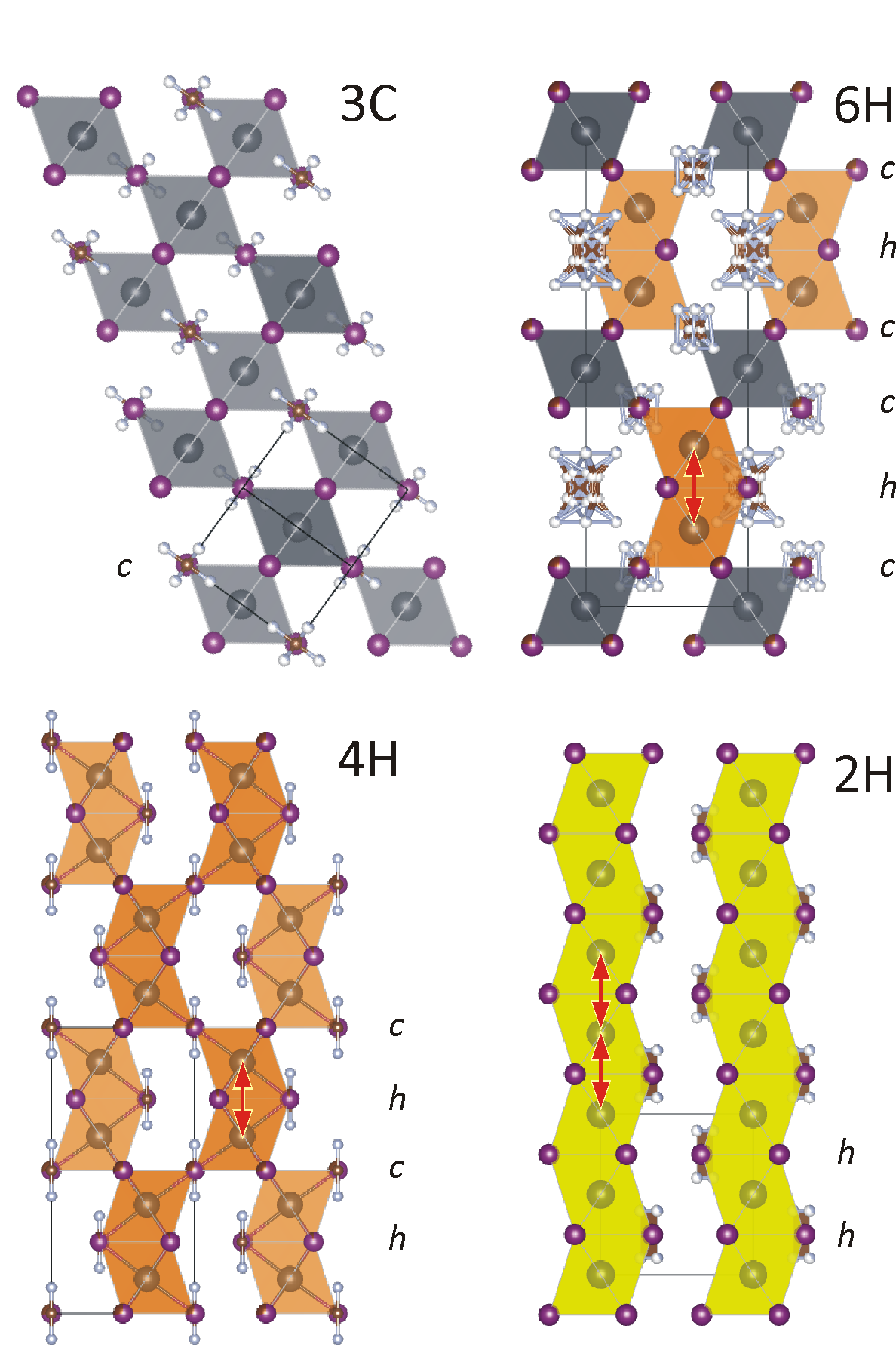}
\caption{{}Structures of the 3C, 6H, 4H and 2H phases of FAPI, based on the
data of Gratia \textit{et al.}\protect\cite{GZS17,Gra18} for FA$_{0.85}$MA$%
_{0.15}$PbI$_{3}$. The red arrows indicate the short Pb--Pb distances across
the shared faces of the octahedra; $c$ and $h$ indicate cubic and hexagonal
stacking of the FAI$_{3}$ layers.}
\label{fig-nH}
\end{figure}

We will base our discussion of the present experimental results on the
hexagonal phases that are intermediate between the well known $\delta $
phase and the cubic $\alpha $ phase and that have been observed in FAPI
partially substituted with MA, Cs and Br.\cite{GZS17,Gra18,SMS19,DWG19} The
structures of these ABX$_{3}$ polymorphs can be described in terms of
stacking of AX$_{3}$ close--packed layers between which the B ions are
octahedrally coordinated with the X ions.\cite{GKL72,LRA90} Adjacent AX$_{3}$
layers are horizontally shifted with respect to each other so that the A
cations are never first neighbors and this gives rise to two types of
stacking. If a layer is sandwiched between identical layers, the BX$_{6}$
octahedra do not touch with each other in the planar directions but share
faces perpendicular to it and form a hexagonal lattice; instead, in a stack
of three different layers the octahedra share corners in all directions and
form the cubic perovskite. Many combinations of cubic and hexagonal stacking
are possible and may be labeled according to Ramsdell's notation,\cite%
{Ram1947} as $n$Y, where $n$ is the number of layers forming the unit cell
and Y\ = C, H, R denotes the cubic, hexagonal, rhombohedral symmetry of the
resulting lattice. The polytypes found in the FA--based halide perovskites
are shown in Fig. \ref{fig-nH} and are: cubic (3C\ = $\alpha $ phase); pure
hexagonal ($h$) stacking (2H\ = $\delta $ phase), where columns of
face-sharing octahedra are separated by FA cations; mixed $hchc$ (4H) and $%
chcchc$ (6H) stacking.

In the perovskite based polytypes, the reason why the 2H phase is more
stable than 3C is that in the latter the A cation is too large for its
cavity surrounded by eight BX$_{6}$ octahedra.\cite{GZ98} In the 2H
structure the problem of the mismatch between B--X and A--X bonds is
completely eliminated by the fact that the B--X bonds are confined within
the columns of octahedra, whose spacing is determined by the A--X bonds. In
the columns of face--sharing octahedra, however, the B--B distances are much
reduced with respect to corner--sharing (red arrows in Fig. \ref{fig-nH}),
with consequent increase of the electrostatic repulsion. The balance between
B--B repulsion and B--X and A--X mismatch makes competitive the intermediate
structures with both corner-- and face--sharing octahedra, and these trends
are confirmed by the distortions of the octahedra and displacements of Pb
from their centers in the structures of (FA$_{0.85}$MA$_{0.15}$)PbI$_{3}$:%
\cite{GZS17} from the Crystallographic
Information Files (CIF) supplied in Ref. \citenum{GZS17} it can be
found that in the 4H and 6H structures Pb$^{2+}$ is displaced away from the
shared face in order to increase the separation from the nearest neighbor Pb$%
^{2+}$ atom and the face is shrunk in order to enhance the electrostatic
screening of the three I$^{-}$ anions. Identical observations can be made in
6H--AzPbBr$_{3}$.\cite{TCQ19}

The fact that pressing the yellow powder of 2H phase induces a change of
color to orange already at 0.2~GPa and the color becomes more intense and
dark with increasing pressure, should be explained in terms of
pressure--induced transformations from the 2H to the 4H and 6H phases.
Indeed, the latter are respectively orange and red in FAPI partially
substituted with MA, Cs and Br,\cite{GZS17,Gra18} and this may be explained
in terms of an increase of the band gap when increasing the octahedral
connectivity from corner-- to edge-- and face--sharing.\cite{KWH17} To our
knowledge, a sequence of pressure--induced transformations of the purely
hexagonal 2H structure into the more corner--sharing 4H and 6H structures
has not been reported so far for hybrid halide perovskites. Yet, this is
what happens in other inorganic perovskites, notably fluorides and chlorides,%
\cite{GKL72} but also oxides such as BaRuO$_{3}$,\cite{GZ98} though at
pressures well above 1~GPa and the sequence of phases is not general.

\begin{table*}[th]
\caption{Volumes per formula unit of the polymorphs of (FAPI)$_{0.85}$(MAPBr)%
$_{0.15}$, according to Ref. \citenum{GZS17}.}
\label{vols}\centering
\begin{tabular}{lllll}
\hline
Ramsdell & 3C & 6H & 4H & 2H \\
$Z$ & 1 & 6 & 4 & 2 \\
$V/Z$ [\AA $^{3}$] & 251 & 253.5 & 255.75 & 256.5 \\
$V/\left( Z~V_{\text{3C}}\right) $ & 1 & 1.009 & 1.019 & 1.022 \\
$a$ [\AA ] & 6.31 & 8.84 & 8.814 & 8.667 \\
$2c/Z$ [\AA ] & 6.31 & 7.48 & 7.604 & 7.908 \\
space group & $Pm\overline{3}m$ & $P6_{3}/mmc$ & $P6_{3}/mmc$ & $P6_{3}/mmc$
\\
color & black & dark red & orange & yellow \\ \hline
&  &  &  &
\end{tabular}%
\end{table*}

The fact that pressure induces transitions to other polymorphs may be put in
direct relationship with their smaller volumes per unit formula but a trend
of the relative volumes of the hexagonal and cubic polymorphs of ABX$_{3}$
compounds cannot be deduced simply from geometrical considerations, since
there are strong variations depending on the ion types. For example, among
the inorganic lead halides it was found that CsCdBr$_{3}$ remains in the 2H
phase up to 23~GPa, almost isotropically squeezed down to 65\% of its
original volume.\cite{LRA90} On the other hand, CsPbI$_{3}$ undergoes the $%
\delta \rightarrow \alpha $ transition at $560-600$~K, but the cubic
perovskite phase has a volume 6.9\% larger than that of the $\delta $ phase,
in spite of a 9\%\ reduction of the volume of the octahedra,\cite{TM08} and
CsSnI$_{3}$ behaves similarly.\cite{CSI12} It must be noted, however, that
those $\delta $ phases are not 2H but orthorhombic with double chains of
edge--sharing octahedra and among the hybrid halides there is a variety of
non--perovskite structures.\cite{SMM17b,GYL19}.

Unfortunately, based on our diffraction data and the literature it is not
trivial to find solid information on the volumes per formula unit of the
various polymorphs of FAPI. Most of our diffraction data were acquired
during fast scans in order to minimize possible effects from degradation of
the compacted powders during the scans in air, and were unsuitable
for Rietveld refinement. Therefore we do not deduce cell volumes of the
various phases from our data but we refer to the studies of Gratia \textit{%
et al.}\cite{GZS17,Gra18} on (FAPbI$_{3}$)$_{1-x}$(MAPbBr$_{3}$)$_{x}$ with $%
x\leq $ $0.2$, which should be indicative also for pure FAPI. Table \ref%
{vols} is based on their cell volumes $V$ with multiplicity $Z$ with respect
to the formula unit. The first row reports the type of layer stacking,
according to Ramsdell's notation,\cite{Ram1947} and the table is sorted in
order of increasing molecular volume $V_{m}=$ $V/Z$ (volume per formula
unit). It results that indeed the molecular volume decreases with increase
of the cubic coordination, following the sequence 2H, 4H, 6H nd 3C. There is
another hybrid halide where the compactness of the 3C phase with respect to
the hexagonal ones is even more marked:\ a stable 6H phase has been found in
Az$_{1-x}$MA$_{x}$PbBr$_{3}$\ (Az = azetidinium = C$_{3}$H$_{6}$NH$_{2}$)$^+$
with a coexistence region $0.4<$\ $x<$\ $0.7$\ of the 3C and 6H phases where
$V_{\mathrm{3C}}$ is a remarkable 9.7\% smaller than $V_{\mathrm{6H}}$.\cite%
{TCQ19}

It seems logical to explain the trend of increasing volume when passing from
the 3C to 2H, 4H, 6H structures of the metal--organic halides in terms of an
overall loss of connectivity of the octahedra, whose strong Pb-X bonds
restrain the lattice expansion due to the large A molecular cations. When
the connectivity is purely one-dimensional along the chains of the octahedra
in the 2H phase, the distance between the chains can more freely accommodate
the molecular cations, in the absence of strong A-A and A-X bonds.
Increasing $n$ in the $n$H series progressively redistributes the octahedral
connectivity from face sharing along $c$ to edge sharing in all directions
and is indeed accompanied by a decrease of the molecular volume. Only based
on this concept one would expect that within the $n$H series shrinking is
larger within the $ab$ plane than along $c$ (it is not possible to extend
the comparison to $a$ of the 3C cubic structure due to the drastic
structural rearrangements, also including rotations of the octahedra by $%
45^{\circ }$ with respect to the principal axes). Contrary to the naive
expectation, the volume shrinking from 2H to 6H is totally due a contraction
along $c$, while there is even a small expansion within the $ab$ plane. This
can be understood in terms of the strong electrostatic repulsion between the
Pb$^{2+}$ atoms on either sides of the 3I$^-$ faces shared by adjacent octahedra, as
explained above.\cite{GZ98,SSR07,DKH18,TCQ19} It seems therefore that the
large size of the FA molecules drives the volume expansion along the 3C to
2H\ series, but the distortion of the cell is driven by the electrostatic
repulsion of the Pb atoms along the columns of face sharing octahedra.

In this context it is usual to introduce the concept of tolerance factor
defined in Eq. (\ref{eq-t}). When $t<$ $1$ the BX$_{6}$ octahedra are too
large to fit the A--X bonds and therefore, if the B--X bonds are more rigid
than the A--X ones, they tilt if an external pressure is applied or when the
temperature is lowered and the thermal shrinking of the more anharmonic A--X
bonds exerts enough pressure.\cite{CCT17} On the opposite instance of $t>$ $1$ the
polymorphs with octahedra sharing faces or edges become favored over cubic
perovskite, as explained above. A particularity of FAPI is that at room
temperature it exhibits both the phenomena typical of perovskites with large
$t$, namely the formation of the hexagonal phases of face--sharing
octahedra, and small $t$, namely tilting of the octahedra just below room
temperature. This is probably due to an important role of the H bonds
between FA and PbI$_{6}$ octahedra,\cite{CCT18} which adds to the usual
steric constraint of fitting cations into the spaces between the octahedra,
and to an exceptionally large thermal expansion,\cite{FSL16} which
accelerates the shrinking of the FA--I network with respect to the Pb--I one
during cooling and causes tilting.

\subsection{Intercalated water as catalyst of the structural transitions}

The comparison between the dielectric measurement of the initially humid
sample in a small closed volume (curve 1 of Fig. \ref{fig-diel}) and the
anelastic measurements in HV reveals that interstitial water molecules
promote the transformations from the hexagonal $n$H phases to the perovskite
3C. This is further confirmed by the anelastic experiment with heating in
humid atmosphere (Fig. \ref{fig-FAPI5-7}).

The interstitial water molecules certainly interfere with the H bonds
between formamidinium molecules and I$^{-}$ anions, which contribute keeping
the $n$H structures, and likely form H bonds with the I$^{-}$ ions, so
weakening the Pb-I bonds and helping the rearrangement process of the PbI$%
_{6}$ octahedra. The role of interstitial water as catalyst has been
demonstrated in MAPI and FAPI for the degradation transformations but may
well exist for the reverse transformations at higher temperature. In
addition, humidity promotes the crystallization of large grains of MAPI and
FAPI films,\cite{HTL17} and accelerates the reaction between MAI and PbI$%
_{2} $ in the mechanosynthesis of MAPI.\cite{MYM16} Similar mechanisms act
in other metal-organic frameworks or coordination polymers.\cite{WL05}
Remaining with the halide perovskites, it has been demonstrated by DSC under
dry and humid conditions that water catalyzes the $\alpha \rightarrow \delta
$ transition in CsPbI$_{3}$;\cite{DHD17} here we see that, at least for
FAPI, also the reverse transition is accelerated. As suggested for CsPbI$%
_{3} $, intercalated water lowers the enthalpy barrier between the
structural variants without affecting their equilibrium enthalpies. Indeed,
in a very recent paper\cite{YDH20} it has been demonstrated that, in the
presence of vapors of solvents such as DMSO, $\delta -$FAPI films transform
fast into large grain $\alpha -$FAPI, a fact attributed to the catalytic
effect of the solvent molecules in promoting the $\delta \rightarrow \alpha $
transition at the phase interface.\cite{YDH20} Water would be a more
effective solvent than the organic ones, thanks to its smaller size and
stronger polarity (the polarity index of water is 10.2 while that of the
most polar of the organic solvents used with FAPI, DMSO, is 7.2\cite{YDH20}%
). Even interstitial inert gases interfere with the structural transitions
in hybrid halide perovskites, as demonstrated by the fact that pressure
induced transitions occur at broadly different pressures depending whether
the pressure transmitting medium is Ne or Ar.\cite{ASK17}

Other observations
are in agreement with a catalytic role of intercalated water: a controlled
amount of humidity ($20-30\%$ RH) improves the morphology of films,\cite%
{YYH14,SM18} though this fact has been mainly attributed to adsorbed H$_{2}$%
O that would help the coalescence of the grains, as in MgO humid sintering.%
\cite{AM64} Similarly, the improved quality of films fabricated in 30\% RH
compared to dry conditions has been attributed to dissolution of the
reactants and enhanced mass transport.\cite{ZCL14} On the other hand, the
acceleration of the kinetics of the structural transitions from the hexagonal to
cubic phase points to a role of H$_{2}$O intercalated within the bulk rather
than in liquid form at the grain boundaries. The latter may rather be
responsible for the degradation to PbI$_{2}$.

It seems therefore worth to more accurately explore the role of small
controlled amounts of water or humidity in the various phases of the powder
or film preparations, in order to find the right balance for a complete
formation of $\alpha $ phase with good microstructure but avoiding the
decomposition into PbI$_{2}$ and organic constituents. Indeed, the
dielectric experiments on FAPI \#4 are made in presence of small
uncontrolled amounts of water and the resulting XRD spectra contain no trace
of PbI$_{2}$ after the first set of measurements (pattern 1 of Fig. \ref{fig-XRD-FAPI4}).
The proof that water was present in the measuring cell is the
complete transformation back to the $\delta $ phase after three days, while
samples of FAPI can be maintained in the $\alpha $ phase for months in the
HV insert.
On the other hand, beyond some threshold water induces decomposition, and
indeed traces of PbI$_2$ were detected in sample \#4 after additional
temperature runs (curve 2 of Fig. \ref{fig-XRD-FAPI4}). Also for FAPI \#7
heated in 7~mbar H$_{2}$O up to 450~K in two hours the subsequent XRD analysis
indicated decomposition, though the amount of PbI$_{2}$
is difficult to asses. The anelastic spectra measured in HV after the
transformation to the $\alpha $ phase showed a perfect perovskite with full
transitions to the $\beta $ and $\gamma $ phases. Yet, the presence of PbI$%
_{2}$ phase probably does not cause any major elastic anomalies and simply
results in a slightly lower elastic modulus (and dielectric constant).
Once extracted from the HV insert, the sample was black with non
uniform yellow shades (Fig. \ref{S2}(b)), and
was cut into pieces to check that also the interior was black. Against the
expectation from the anelastic and visual evaluation, the XRD spectrum
measured in reflection from the sample surface presented a prevalence of the
PbI$_{2}$ peaks over the $\alpha -$FAPI ones. As already noted in
reference to Fig. \ref{fig-XRD-FAPI4}, we attribute this fact
to a superior crystallinity of the hexagonal phases over cubic $\alpha -$FAPI.
Indeed,
the transformation of $\alpha -$FAPI into $\delta $ phase or its decomposition
into PbI$_2$ catalyzed or induced by water at room temperature
occur relatively slowly, presumably from few nuclei with larger concentration
of H$_2$O molecules, so allowing large hexagonal crystallites to grow.\cite{CDW17}
On the contrary, the transformation from the $\delta $ to the cubic phase
occurs at high temperature with a faster kinetics,
especially if catalyzed by intercalated water.
As explained above, the transitions from the $n$H phases to 3C do
not require long range ion diffusion but only rearrangements in the stacking
of the PbI$_{3}$ planes. If this occurs simultaneously at closely spaced
intervals, nanometric domain configurations may arise, which indeed have
been observed in $\alpha -$FAPI and mimic a trigonal or hexagonal
symmetry rather than cubic.\cite{WWF15,WGG18} It is then possible that
the X-ray reflections from the so nanotwinned cubic phase are much less
sharp than those from nearly perfect crystallites of PbI$_{2}$
and $\delta $ phase.

\section{Conclusions}

We studied the phase transitions between the various polymorphs of FAPI by
anelastic, dielectric and X-ray diffraction measurements on samples pressed
from $\delta -$FAPI (2H phase) yellow powder. The application of a pressure
as low as 0.2~GPa changes the color to orange, which has been explained in
terms of transformations to the other hexagonal polymorphs 4H and 6H, having
smaller volumes. The stability of these intermediate hexagonal phases has
been put in the broader perspective of the hybrid and inorganic perovskites
ABX$_{3}$ with large A cations, or large tolerance factor. It appears that
the several polymorphs of FAPI are rather close in free energy at room
temperature and, remarkably, FAPI is susceptible to both the phenomena due
to a large tolerance factor, namely the formation of the hexagonal phases of
face--sharing octahedra, and small tolerance factor, with octahedral tilting
just below room temperature. This structural susceptibility is certainly
enhanced by the extremely large thermal expansion and is probably a cause
for the contrasting observations on the effects of pressure, temperature and
aging on the FAPI structure.

In addition, it is observed that a small amount of water may enhance the
kinetics of the transformations from the various polytypes to the cubic
phase. Rather than to a surface effect of adsorbed H$_{2}$O, as often
assumed for the improved grain growth in films in moderately humid
atmosphere, this effect is attributed to H$_{2}$O intercalated in the bulk.
The catalytic effect of water presents a similarity with the accomplishment
of fast formation of $\alpha -$FAPI with large grains from $\delta -$FAPI
films in presence of DMSO and other solvents,\cite{YDH20} but the highly
polar H$_{2}$O should be an even more effective solvent. This fact acquires
particular interest in the light of the recent observation that the trapping
of photocarriers, a major cause of reduction of the photovoltaic
performance, occurs at specific types of grain boundaries, so that \
"managing structure and composition on the nanoscale will be essential for
optimal performance of halide perovskite devices."\cite{DWM20} It seems
therefore promising to explore strategies where controlled small amounts of
water are introduced during the synthesis in order to improve the atomic
diffusion and crystallization, and thoroughly removed after a sufficiently
homogeneous and smooth microstructure is achieved.

\begin{acknowledgement}

The authors thank Paolo Massimiliano Latino and Roberto Scaccia (ISM-Tor Vergata) for
their technical assistance in the anelastic experiments, Sara Notarantonio
(ISM-Montelibretti) for assistance in the synthesis, Marco Guaragno
(ISM-Tor Vergata) for his technical support in X-ray experiments.

\end{acknowledgement}


\bibliographystyle{plain}
\bibliography{refs}

\providecommand{\latin}[1]{#1}
\makeatletter
\providecommand{\doi}
  {\begingroup\let\do\@makeother\dospecials
  \catcode`\{=1 \catcode`\}=2 \doi@aux}
\providecommand{\doi@aux}[1]{\endgroup\texttt{#1}}
\makeatother
\providecommand*\mcitethebibliography{\thebibliography}
\csname @ifundefined\endcsname{endmcitethebibliography}
  {\let\endmcitethebibliography\endthebibliography}{}
\begin{mcitethebibliography}{71}
\providecommand*\natexlab[1]{#1}
\providecommand*\mciteSetBstSublistMode[1]{}
\providecommand*\mciteSetBstMaxWidthForm[2]{}
\providecommand*\mciteBstWouldAddEndPuncttrue
  {\def\EndOfBibitem{\unskip.}}
\providecommand*\mciteBstWouldAddEndPunctfalse
  {\let\EndOfBibitem\relax}
\providecommand*\mciteSetBstMidEndSepPunct[3]{}
\providecommand*\mciteSetBstSublistLabelBeginEnd[3]{}
\providecommand*\EndOfBibitem{}
\mciteSetBstSublistMode{f}
\mciteSetBstMaxWidthForm{subitem}{(\alph{mcitesubitemcount})}
\mciteSetBstSublistLabelBeginEnd
  {\mcitemaxwidthsubitemform\space}
  {\relax}
  {\relax}

\bibitem[Green \latin{et~al.}(2014)Green, Ho-Baillie, and Snaith]{GHS14}
Green,~M.~A.; Ho-Baillie,~A.; Snaith,~H.~J. {The emergence of perovskite solar
  cells}. \emph{Nat. Phot.} \textbf{2014}, \emph{8}, 506\relax
\mciteBstWouldAddEndPuncttrue
\mciteSetBstMidEndSepPunct{\mcitedefaultmidpunct}
{\mcitedefaultendpunct}{\mcitedefaultseppunct}\relax
\EndOfBibitem
\bibitem[Correa-Baena \latin{et~al.}(2017)Correa-Baena, Saliba, Buonassisi,
  Gr{\"a}tzel, Abate, Tress, and Hagfeldt]{CSB17}
Correa-Baena,~J.~P.; Saliba,~M.; Buonassisi,~T.; Gr{\"a}tzel,~M.; Abate,~A.;
  Tress,~W.; Hagfeldt,~A. {Promises and challenges of perovskite solar cells}.
  \emph{Science} \textbf{2017}, \emph{358}, 739\relax
\mciteBstWouldAddEndPuncttrue
\mciteSetBstMidEndSepPunct{\mcitedefaultmidpunct}
{\mcitedefaultendpunct}{\mcitedefaultseppunct}\relax
\EndOfBibitem
\bibitem[Sutherland and Sargent(2016)Sutherland, and Sargent]{SS16b}
Sutherland,~B.~R.; Sargent,~E.~H. {Perovskite photonic sources}. \emph{Nat.
  Photonics} \textbf{2016}, \emph{10}, 295\relax
\mciteBstWouldAddEndPuncttrue
\mciteSetBstMidEndSepPunct{\mcitedefaultmidpunct}
{\mcitedefaultendpunct}{\mcitedefaultseppunct}\relax
\EndOfBibitem
\bibitem[Ava \latin{et~al.}(2019)Ava, Mamun, Marsillac, and Namkoong]{AMM19}
Ava,~T.~T.; Mamun,~A.~A.; Marsillac,~S.; Namkoong,~G. {A Review: Thermal
  Stability of Methylammonium Lead Halide Based Perovskite Solar Cells}.
  \emph{Appl. Sci.} \textbf{2019}, \emph{9}, 188\relax
\mciteBstWouldAddEndPuncttrue
\mciteSetBstMidEndSepPunct{\mcitedefaultmidpunct}
{\mcitedefaultendpunct}{\mcitedefaultseppunct}\relax
\EndOfBibitem
\bibitem[Boyd \latin{et~al.}(2019)Boyd, Cheacharoen, Leijtens, and
  McGehee]{BCL19}
Boyd,~C.~C.; Cheacharoen,~R.; Leijtens,~T.; McGehee,~M.~D. {Understanding
  Degradation Mechanisms and Improving Stability of Perovskite Photovoltaics}.
  \emph{Chem. Rev.} \textbf{2019}, \emph{119}, 3418\relax
\mciteBstWouldAddEndPuncttrue
\mciteSetBstMidEndSepPunct{\mcitedefaultmidpunct}
{\mcitedefaultendpunct}{\mcitedefaultseppunct}\relax
\EndOfBibitem
\bibitem[Liu \latin{et~al.}(2019)Liu, Kong, Yang, and Mao]{LKY19}
Liu,~G.; Kong,~L.; Yang,~W.; Mao,~H. {Pressure engineering of photovoltaic
  perovskites}. \emph{Materials Today} \textbf{2019}, \emph{27}, 91\relax
\mciteBstWouldAddEndPuncttrue
\mciteSetBstMidEndSepPunct{\mcitedefaultmidpunct}
{\mcitedefaultendpunct}{\mcitedefaultseppunct}\relax
\EndOfBibitem
\bibitem[Li \latin{et~al.}(2020)Li, Liu, Wang, Yang, and Lu]{LLW20}
Li,~M.; Liu,~T.; Wang,~Y.; Yang,~W.; Lu,~X. {Pressure responses of halide
  perovskites with various compositions, dimensionalities, and morphologie}.
  \emph{Matter Radiat. Extremes} \textbf{2020}, \emph{5}, 018201\relax
\mciteBstWouldAddEndPuncttrue
\mciteSetBstMidEndSepPunct{\mcitedefaultmidpunct}
{\mcitedefaultendpunct}{\mcitedefaultseppunct}\relax
\EndOfBibitem
\bibitem[Kong \latin{et~al.}(2016)Kong, Liu, Gong, Hu, Schaller, Dera, Zhang,
  Liu, Yang, Zhu, Tang, Wang, Wei, Xu, and Mao]{KLG16}
Kong,~L.; Liu,~G.; Gong,~J.; Hu,~Q.; Schaller,~R.~D.; Dera,~P.; Zhang,~D.;
  Liu,~Z.; Yang,~W.; Zhu,~K. \latin{et~al.}  {Simultaneous band-gap narrowing
  and carrier-lifetime prolongation of organic-inorganic trihalide
  perovskites}. \emph{PNAS} \textbf{2016}, \emph{113}, 8910\relax
\mciteBstWouldAddEndPuncttrue
\mciteSetBstMidEndSepPunct{\mcitedefaultmidpunct}
{\mcitedefaultendpunct}{\mcitedefaultseppunct}\relax
\EndOfBibitem
\bibitem[Ghosh \latin{et~al.}(2019)Ghosh, Aziz, Dawson, Walker, and
  Islam]{GAD19}
Ghosh,~D.; Aziz,~A.; Dawson,~J.~A.; Walker,~A.~B.; Islam,~M.~S. {Putting the
  Squeeze on Lead Iodide Perovskites: Pressure-Induced Effects To Tune Their
  Structural and Optoelectronic Behavior}. \emph{Chem. Mater.} \textbf{2019},
  \emph{31}, 4063\relax
\mciteBstWouldAddEndPuncttrue
\mciteSetBstMidEndSepPunct{\mcitedefaultmidpunct}
{\mcitedefaultendpunct}{\mcitedefaultseppunct}\relax
\EndOfBibitem
\bibitem[Goldschmidt(1926)]{Gol26}
Goldschmidt,~V.~M. {Die Gesetze der Krystallochemie}.
  \emph{Naturwissenschaften} \textbf{1926}, \emph{14}, 477\relax
\mciteBstWouldAddEndPuncttrue
\mciteSetBstMidEndSepPunct{\mcitedefaultmidpunct}
{\mcitedefaultendpunct}{\mcitedefaultseppunct}\relax
\EndOfBibitem
\bibitem[Megaw(1952)]{Meg52}
Megaw,~H.~D. Origin of Ferroelectricity in Barium Titanate and Other
  Perovskite-Type Crystals. \emph{Acta Cryst.} \textbf{1952}, \emph{5},
  739\relax
\mciteBstWouldAddEndPuncttrue
\mciteSetBstMidEndSepPunct{\mcitedefaultmidpunct}
{\mcitedefaultendpunct}{\mcitedefaultseppunct}\relax
\EndOfBibitem
\bibitem[Lee \latin{et~al.}(2016)Lee, Bristowe, Lee, Lee, Bristowe, Cheetham,
  and Jang]{LBL16}
Lee,~J.~H.; Bristowe,~N.~C.; Lee,~J.~H.; Lee,~S.~H.; Bristowe,~P.~D.;
  Cheetham,~A.~K.; Jang,~H.~M. {Resolving the Physical Origin of Octahedral
  Tilting in Halide Perovskites}. \emph{Chem. Mater.} \textbf{2016}, \emph{28},
  4259\relax
\mciteBstWouldAddEndPuncttrue
\mciteSetBstMidEndSepPunct{\mcitedefaultmidpunct}
{\mcitedefaultendpunct}{\mcitedefaultseppunct}\relax
\EndOfBibitem
\bibitem[Stoumpos \latin{et~al.}(2013)Stoumpos, Malliakas, and
  Kanatzidis]{SMK13c}
Stoumpos,~C.~C.; Malliakas,~C.~D.; Kanatzidis,~M.~G. {Semiconducting Tin and
  Lead Iodide Perovskites with Organic Cations: Phase Transitions, High
  Mobilities, and Near-Infrared Photoluminescent Properties}. \emph{Inorg.
  Chem.} \textbf{2013}, \emph{52}, 9019\relax
\mciteBstWouldAddEndPuncttrue
\mciteSetBstMidEndSepPunct{\mcitedefaultmidpunct}
{\mcitedefaultendpunct}{\mcitedefaultseppunct}\relax
\EndOfBibitem
\bibitem[Stoumpos and Kanatzidis(2015)Stoumpos, and Kanatzidis]{SK15}
Stoumpos,~C.~C.; Kanatzidis,~M.~G. {The Renaissance of Halide Perovskites and
  Their Evolution as Emerging Semiconductors}. \emph{Acc. Chem. Res.}
  \textbf{2015}, \emph{48}, 2791\relax
\mciteBstWouldAddEndPuncttrue
\mciteSetBstMidEndSepPunct{\mcitedefaultmidpunct}
{\mcitedefaultendpunct}{\mcitedefaultseppunct}\relax
\EndOfBibitem
\bibitem[Gratia \latin{et~al.}(2017)Gratia, Zimmermann, Schouwink, Yum,
  Audinot, Sivula, Wirtz, and Nazeeruddin]{GZS17}
Gratia,~P.; Zimmermann,~I.; Schouwink,~P.; Yum,~J.~H.; Audinot,~J.~N.;
  Sivula,~K.; Wirtz,~T.; Nazeeruddin,~M.~K. {The Many Faces of Mixed Ion
  Perovskites: Unraveling and Understanding the Crystallization Process}.
  \emph{ACS Energy Lett.} \textbf{2017}, \emph{2}, 2686\relax
\mciteBstWouldAddEndPuncttrue
\mciteSetBstMidEndSepPunct{\mcitedefaultmidpunct}
{\mcitedefaultendpunct}{\mcitedefaultseppunct}\relax
\EndOfBibitem
\bibitem[Gratia(2018)]{Gra18}
Gratia,~P. {Compositional Characterization of Organo-Lead tri-Halide Perovskite
  Solar Cells}. Ph.D.\ thesis, 2018\relax
\mciteBstWouldAddEndPuncttrue
\mciteSetBstMidEndSepPunct{\mcitedefaultmidpunct}
{\mcitedefaultendpunct}{\mcitedefaultseppunct}\relax
\EndOfBibitem
\bibitem[Szostak \latin{et~al.}(2019)Szostak, Marchezi, dos Santos~Marques,
  da~Silva, de~Holanda, Soares, Tolentino, and Nogueira]{SMS19}
Szostak,~R.; Marchezi,~P.~E.; dos Santos~Marques,~A.; da~Silva,~J.~C.;
  de~Holanda,~M.~S.; Soares,~M.~M.; Tolentino,~H. C.~N.; Nogueira,~A.~F.
  {Exploring the formation of formamidinium-based hybrid perovskites by
  antisolvent methods: in situ GIWAXS measurements during spin coating}.
  \emph{Sustainable Energy Fuels} \textbf{2019}, \emph{3}, 2287\relax
\mciteBstWouldAddEndPuncttrue
\mciteSetBstMidEndSepPunct{\mcitedefaultmidpunct}
{\mcitedefaultendpunct}{\mcitedefaultseppunct}\relax
\EndOfBibitem
\bibitem[Dang \latin{et~al.}(2019)Dang, Wang, Ghasemi, Tang, Bastiani, Aydin,
  Dauzon, Barrit, Peng, Smilgies, Wolf, and Amassian]{DWG19}
Dang,~H.~X.; Wang,~K.; Ghasemi,~M.; Tang,~M.~C.; Bastiani,~M.~D.; Aydin,~E.;
  Dauzon,~E.; Barrit,~D.; Peng,~J.; Smilgies,~D.~M. \latin{et~al.}
  {Multi-cation Synergy Suppresses Phase Segregation in Mixed-Halide
  Perovskites}. \emph{Joule} \textbf{2019}, \emph{3}, 1746\relax
\mciteBstWouldAddEndPuncttrue
\mciteSetBstMidEndSepPunct{\mcitedefaultmidpunct}
{\mcitedefaultendpunct}{\mcitedefaultseppunct}\relax
\EndOfBibitem
\bibitem[Ramsdell(1947)]{Ram1947}
Ramsdell,~L.~S. {Studies on silicon carbide}. \emph{Am. Mineral.}
  \textbf{1947}, \emph{32}, 64\relax
\mciteBstWouldAddEndPuncttrue
\mciteSetBstMidEndSepPunct{\mcitedefaultmidpunct}
{\mcitedefaultendpunct}{\mcitedefaultseppunct}\relax
\EndOfBibitem
\bibitem[Ding \latin{et~al.}(2018)Ding, Li, Li, Li, Yao, Liu, Tian, Su, Chen,
  and Shi]{DLL18}
Ding,~D.; Li,~H.; Li,~J.; Li,~Z.; Yao,~H.; Liu,~L.; Tian,~B.~B.; Su,~C.;
  Chen,~F.; Shi,~Y. {Effect of mechanical forces on thermal stability
  reinforcement for lead based perovskite materials}. \emph{J. Mater. Chem. A}
  \textbf{2018}, \emph{7}, 540\relax
\mciteBstWouldAddEndPuncttrue
\mciteSetBstMidEndSepPunct{\mcitedefaultmidpunct}
{\mcitedefaultendpunct}{\mcitedefaultseppunct}\relax
\EndOfBibitem
\bibitem[Hong \latin{et~al.}(2019)Hong, Tan, John, Kang, Tay, Ho, Zhao, Sum,
  Mathews, Garcia, and Soo]{HTJ19}
Hong,~Z.; Tan,~D.; John,~R.~A.; Kang,~Y.; Tay,~E.; Ho,~Y. K.~T.; Zhao,~X.;
  Sum,~T.~C.; Mathews,~N.; Garcia,~F. \latin{et~al.}  {Completely Solvent-free
  Protocols to Access Phase-Pure, Metastable Metal Halide Perovskites and
  Functional Photodetectors from the Precursor Salts}. \emph{iScience}
  \textbf{2019}, \emph{16}, 312\relax
\mciteBstWouldAddEndPuncttrue
\mciteSetBstMidEndSepPunct{\mcitedefaultmidpunct}
{\mcitedefaultendpunct}{\mcitedefaultseppunct}\relax
\EndOfBibitem
\bibitem[Manukyan \latin{et~al.}(2016)Manukyan, Yeghishyan, Moskovskikh,
  Kapaldo, Mintairov, and Mukasyan]{MYM16}
Manukyan,~K.~V.; Yeghishyan,~A.~V.; Moskovskikh,~D.~O.; Kapaldo,~J.;
  Mintairov,~A.; Mukasyan,~A.~S. {Mechanochemical synthesis of methylammonium
  lead iodide perovskite}. \emph{J. Mater. Sci .} \textbf{2016}, \emph{51},
  9123\relax
\mciteBstWouldAddEndPuncttrue
\mciteSetBstMidEndSepPunct{\mcitedefaultmidpunct}
{\mcitedefaultendpunct}{\mcitedefaultseppunct}\relax
\EndOfBibitem
\bibitem[Luo \latin{et~al.}(2016)Luo, Ran, Chen, Shen, and Zhang]{LRC16}
Luo,~Y.; Ran,~G.; Chen,~N.; Shen,~Q.; Zhang,~Y. {Microstructural Evolution,
  Thermodynamics, and Kinetics of Mo-Tm$_2$O$_3$ Powder Mixtures during Ball
  Milling}. \emph{Materials} \textbf{2016}, \emph{9}, 834\relax
\mciteBstWouldAddEndPuncttrue
\mciteSetBstMidEndSepPunct{\mcitedefaultmidpunct}
{\mcitedefaultendpunct}{\mcitedefaultseppunct}\relax
\EndOfBibitem
\bibitem[L{\"u} \latin{et~al.}(2016)L{\"u}, Wang, Stoumpos, Hu, Guo, Chen,
  Yang, Smith, Yang, Zhao, Xu, Kanatzidis, and Jia]{LWS16}
L{\"u},~X.; Wang,~Y.; Stoumpos,~C.~C.; Hu,~Q.; Guo,~X.; Chen,~H.; Yang,~L.;
  Smith,~J.~S.; Yang,~W.; Zhao,~Y. \latin{et~al.}  {Enhanced Structural
  Stability and Photo Responsiveness of CH3NH3SnI3 Perovskite via
  Pressure-Induced Amorphization and Recrystallization}. \emph{Adv. Mater.}
  \textbf{2016}, \emph{28}, 8663\relax
\mciteBstWouldAddEndPuncttrue
\mciteSetBstMidEndSepPunct{\mcitedefaultmidpunct}
{\mcitedefaultendpunct}{\mcitedefaultseppunct}\relax
\EndOfBibitem
\bibitem[Liu \latin{et~al.}(2018)Liu, Gong, Kong, Schaller, Hu, Liu, Yan, Yang,
  Stoumpos, Kanatzidis, Mao, and Xu]{LGK18}
Liu,~G.; Gong,~J.; Kong,~L.; Schaller,~R.~D.; Hu,~Q.; Liu,~Z.; Yan,~S.;
  Yang,~W.; Stoumpos,~C.~C.; Kanatzidis,~M.~G. \latin{et~al.}  {Isothermal
  pressure-derived metastable states in 2D hybrid perovskites showing enduring
  bandgap narrowing}. \emph{PNAS} \textbf{2018}, \emph{115}, 8076\relax
\mciteBstWouldAddEndPuncttrue
\mciteSetBstMidEndSepPunct{\mcitedefaultmidpunct}
{\mcitedefaultendpunct}{\mcitedefaultseppunct}\relax
\EndOfBibitem
\bibitem[Liu \latin{et~al.}(2017)Liu, Kong, Gong, Yang, Mao, Hu, Liu, Schaller,
  Zhang, and Xu]{LKG17}
Liu,~G.; Kong,~L.; Gong,~J.; Yang,~W.; Mao,~H.; Hu,~Q.; Liu,~Z.;
  Schaller,~R.~D.; Zhang,~D.; Xu,~T. {Pressure-Induced Bandgap Optimization in
  Lead-Based Perovskites with Prolonged Carrier Lifetime and Ambient
  Retainability}. \emph{Adv. Funct. Mater.} \textbf{2017}, \emph{27},
  1604208\relax
\mciteBstWouldAddEndPuncttrue
\mciteSetBstMidEndSepPunct{\mcitedefaultmidpunct}
{\mcitedefaultendpunct}{\mcitedefaultseppunct}\relax
\EndOfBibitem
\bibitem[Postorino and Malavasi(2017)Postorino, and Malavasi]{PM17}
Postorino,~P.; Malavasi,~L. {Pressure-Induced Effects in Organic-Inorganic
  Hybrid Perovskites}. \emph{J. Phys. Chem. Lett.} \textbf{2017}, \emph{8},
  2613\relax
\mciteBstWouldAddEndPuncttrue
\mciteSetBstMidEndSepPunct{\mcitedefaultmidpunct}
{\mcitedefaultendpunct}{\mcitedefaultseppunct}\relax
\EndOfBibitem
\bibitem[Jiang \latin{et~al.}(2018)Jiang, Luan, Jang, Baikie, Huang, Li,
  Saouma, Wang, White, and Fang]{JLJ18}
Jiang,~S.; Luan,~Y.; Jang,~J.~I.; Baikie,~T.; Huang,~X.; Li,~R.; Saouma,~F.~O.;
  Wang,~Z.; White,~T.~J.; Fang,~J. {Phase Transitions of Formamidinium Lead
  Iodide Perovskite under Pressure}. \emph{J. Am. Chem. Soc.} \textbf{2018},
  \emph{140}, 13952\relax
\mciteBstWouldAddEndPuncttrue
\mciteSetBstMidEndSepPunct{\mcitedefaultmidpunct}
{\mcitedefaultendpunct}{\mcitedefaultseppunct}\relax
\EndOfBibitem
\bibitem[Sun \latin{et~al.}(2017)Sun, Deng, Wu, Wei, Isikgor, Brivio, Gaultois,
  Ouyang, Bristowe, Cheetham, and Kieslich]{SDW17}
Sun,~S.; Deng,~Z.; Wu,~Y.; Wei,~F.; Isikgor,~F.~H.; Brivio,~F.;
  Gaultois,~M.~W.; Ouyang,~J.; Bristowe,~P.~D.; Cheetham,~A.~K. \latin{et~al.}
  {Variable Temperature and High-Pressure Crystal Chemistry of Perovskite
  Formamidinium Lead Iodide: a Single Crystal X-Ray Diffraction and
  Computational Study}. \emph{Chem. Commun.} \textbf{2017}, \emph{53},
  7537\relax
\mciteBstWouldAddEndPuncttrue
\mciteSetBstMidEndSepPunct{\mcitedefaultmidpunct}
{\mcitedefaultendpunct}{\mcitedefaultseppunct}\relax
\EndOfBibitem
\bibitem[Wang \latin{et~al.}(2017)Wang, Guan, Galeschuk, Yao, He, Jiang, Zhang,
  Liu, Jin, Jin, and Song]{WGG17}
Wang,~P.; Guan,~J.; Galeschuk,~D.~K.; Yao,~Y.; He,~C.~F.; Jiang,~S.; Zhang,~S.;
  Liu,~Y.; Jin,~M.; Jin,~C. \latin{et~al.}  {Pressure-Induced Polymorphic,
  Optical, and Electronic Transitions of Formamidinium Lead Iodide Perovskite}.
  \emph{J. Phys. Chem. Lett.} \textbf{2017}, \emph{89}, 2119\relax
\mciteBstWouldAddEndPuncttrue
\mciteSetBstMidEndSepPunct{\mcitedefaultmidpunct}
{\mcitedefaultendpunct}{\mcitedefaultseppunct}\relax
\EndOfBibitem
\bibitem[Zhu \latin{et~al.}(2018)Zhu, Cai, Que, Song, Rubenstein, Wang, and
  Chen]{ZCQ18}
Zhu,~H.; Cai,~T.; Que,~M.; Song,~J.~P.; Rubenstein,~B.~M.; Wang,~Z.; Chen,~O.
  {Pressure-Induced Phase Transformation and Band-Gap Engineering of
  Formamidinium Lead Iodide Perovskite Nanocrystals}. \emph{J. Phys. Chem.
  Lett.} \textbf{2018}, \emph{9}, 4199\relax
\mciteBstWouldAddEndPuncttrue
\mciteSetBstMidEndSepPunct{\mcitedefaultmidpunct}
{\mcitedefaultendpunct}{\mcitedefaultseppunct}\relax
\EndOfBibitem
\bibitem[Wang \latin{et~al.}(2016)Wang, Wang, and Zou]{WWZ16}
Wang,~L.; Wang,~K.; Zou,~B. {Pressure-Induced Structural and Optical Properties
  of Organometal Halide Perovskite-Based Formamidinium Lead Bromide}. \emph{J.
  Phys. Chem. Lett.} \textbf{2016}, \emph{7}, 2556\relax
\mciteBstWouldAddEndPuncttrue
\mciteSetBstMidEndSepPunct{\mcitedefaultmidpunct}
{\mcitedefaultendpunct}{\mcitedefaultseppunct}\relax
\EndOfBibitem
\bibitem[Zhou \latin{et~al.}(2014)Zhou, Chen, Li, Luo, Song, Duan, Hong, You,
  Liu, and Yang]{ZCL14}
Zhou,~H.; Chen,~Q.; Li,~G.; Luo,~S.; Song,~T.; Duan,~H.~S.; Hong,~Z.; You,~J.;
  Liu,~Y.; Yang,~Y. {Interface engineering of highly efficient perovskite solar
  cells}. \emph{Science} \textbf{2014}, \emph{345}, 542\relax
\mciteBstWouldAddEndPuncttrue
\mciteSetBstMidEndSepPunct{\mcitedefaultmidpunct}
{\mcitedefaultendpunct}{\mcitedefaultseppunct}\relax
\EndOfBibitem
\bibitem[Huang \latin{et~al.}(2017)Huang, Tan, Lund, and Zhou]{HTL17}
Huang,~J.; Tan,~S.; Lund,~P.~D.; Zhou,~H. {Impact of H2O on organic-inorganic
  hybrid perovskite solar cells}. \emph{Energy Environ. Sci.} \textbf{2017},
  \emph{10}, 2284\relax
\mciteBstWouldAddEndPuncttrue
\mciteSetBstMidEndSepPunct{\mcitedefaultmidpunct}
{\mcitedefaultendpunct}{\mcitedefaultseppunct}\relax
\EndOfBibitem
\bibitem[Liu \latin{et~al.}(2016)Liu, Sun, Yang, Yang, Ren, Xu, Yang, and
  Liu]{LSY16}
Liu,~Y.; Sun,~J.; Yang,~Z.; Yang,~D.; Ren,~X.; Xu,~H.; Yang,~Z.; Liu,~S.~F.
  {20-mm-Large Single-Crystalline Formamidinium-Perovskite Wafer for Mass
  Production of Integrated Photodetectors}. \emph{Adv. Opt. Mater.}
  \textbf{2016}, \emph{4}, 1829\relax
\mciteBstWouldAddEndPuncttrue
\mciteSetBstMidEndSepPunct{\mcitedefaultmidpunct}
{\mcitedefaultendpunct}{\mcitedefaultseppunct}\relax
\EndOfBibitem
\bibitem[Jana \latin{et~al.}(2019)Jana, Ba, Nissimagoudar, and Kim]{JBN19}
Jana,~A.; Ba,~Q.; Nissimagoudar,~A.~S.; Kim,~K.~S. {Formation of a photoactive
  quasi-2D formamidinium lead iodide perovskite in water}. \emph{J. Mater.
  Chem. A} \textbf{2019}, \emph{7}, 25785\relax
\mciteBstWouldAddEndPuncttrue
\mciteSetBstMidEndSepPunct{\mcitedefaultmidpunct}
{\mcitedefaultendpunct}{\mcitedefaultseppunct}\relax
\EndOfBibitem
\bibitem[Cordero \latin{et~al.}(2009)Cordero, Bella, Corvasce, Latino, and
  Morbidini]{CDC09}
Cordero,~F.; Bella,~L.~D.; Corvasce,~F.; Latino,~P.~M.; Morbidini,~A. {An
  insert for anelastic spectroscopy measurements from 80~K to 1100~K}.
  \emph{Meas. Sci. Technol.} \textbf{2009}, \emph{20}, 015702\relax
\mciteBstWouldAddEndPuncttrue
\mciteSetBstMidEndSepPunct{\mcitedefaultmidpunct}
{\mcitedefaultendpunct}{\mcitedefaultseppunct}\relax
\EndOfBibitem
\bibitem[Nowick and Berry(1972)Nowick, and Berry]{NB72}
Nowick,~A.~S.; Berry,~B.~S. \emph{Anelastic Relaxation in Crystalline Solids};
  Academic Press: New York, 1972\relax
\mciteBstWouldAddEndPuncttrue
\mciteSetBstMidEndSepPunct{\mcitedefaultmidpunct}
{\mcitedefaultendpunct}{\mcitedefaultseppunct}\relax
\EndOfBibitem
\bibitem[Cordero \latin{et~al.}(2019)Cordero, Craciun, Trequattrini, Generosi,
  Paci, Paoletti, and Pennesi]{CCT19}
Cordero,~F.; Craciun,~F.; Trequattrini,~F.; Generosi,~A.; Paci,~B.;
  Paoletti,~A.~M.; Pennesi,~G. {Stability of Cubic FAPbI$_3$ from X-ray
  Diffraction, Anelastic, and Dielectric Measurements}. \emph{J. Phys. Chem.
  Lett.} \textbf{2019}, \emph{10}, 2463\relax
\mciteBstWouldAddEndPuncttrue
\mciteSetBstMidEndSepPunct{\mcitedefaultmidpunct}
{\mcitedefaultendpunct}{\mcitedefaultseppunct}\relax
\EndOfBibitem
\bibitem[Keshavarz \latin{et~al.}(2019)Keshavarz, Ottesen, Wiedmann, Wharmby,
  Yuan, Debroye, Steele, Martens, Hussey, Roeffaers, and Hofkens]{KOW19}
Keshavarz,~M.; Ottesen,~M.; Wiedmann,~S.; Wharmby,~M.; Yuan,~R. K.~H.;
  Debroye,~E.; Steele,~J.~A.; Martens,~J.; Hussey,~N.~E.; Roeffaers,~M. B.
  M.~J. \latin{et~al.}  {Tracking Structural Phase Transitions in Lead-Halide
  Perovskites by Means of Thermal Expansion}. \emph{Adv. Mater.} \textbf{2019},
  1900521\relax
\mciteBstWouldAddEndPuncttrue
\mciteSetBstMidEndSepPunct{\mcitedefaultmidpunct}
{\mcitedefaultendpunct}{\mcitedefaultseppunct}\relax
\EndOfBibitem
\bibitem[Chen \latin{et~al.}(2016)Chen, Foley, Park, Brown, Harriger, Lee,
  Ruff, Yoon, Choi, and Lee]{CFP16}
Chen,~T.; Foley,~B.~J.; Park,~C.; Brown,~C.~M.; Harriger,~L.~W.; Lee,~J.;
  Ruff,~J.; Yoon,~M.; Choi,~J.~J.; Lee,~S.~H. {Entropy-Driven Structural
  Transition and Kinetic Trapping in Formamidinium Lead Iodide Perovskite}.
  \emph{Sci. Adv.} \textbf{2016}, \emph{2}, e1601650\relax
\mciteBstWouldAddEndPuncttrue
\mciteSetBstMidEndSepPunct{\mcitedefaultmidpunct}
{\mcitedefaultendpunct}{\mcitedefaultseppunct}\relax
\EndOfBibitem
\bibitem[Chen \latin{et~al.}(2017)Chen, Chen, Foley, Lee, Ruff, Ko, Brown,
  Harriger, Zhang, Park, Yoon, Chang, Choi, and Lee]{CCF17c}
Chen,~T.; Chen,~W.~L.; Foley,~B.~J.; Lee,~J.; Ruff,~J.~C.; Ko,~J. Y.~P.;
  Brown,~C.~M.; Harriger,~L.~W.; Zhang,~D.; Park,~C. \latin{et~al.}  {Origin of
  Long Lifetime of Band-Edge Charge Carriers in Organic-Inorganic Lead Iodide
  Perovskites}. \emph{PNAS} \textbf{2017}, \emph{114}, 7519\relax
\mciteBstWouldAddEndPuncttrue
\mciteSetBstMidEndSepPunct{\mcitedefaultmidpunct}
{\mcitedefaultendpunct}{\mcitedefaultseppunct}\relax
\EndOfBibitem
\bibitem[Wadhawan(1982)]{Wad82}
Wadhawan,~V.~K. Ferroelasticity and related properties of crystals. \emph{Phase
  Transitions} \textbf{1982}, \emph{3}, 3--103\relax
\mciteBstWouldAddEndPuncttrue
\mciteSetBstMidEndSepPunct{\mcitedefaultmidpunct}
{\mcitedefaultendpunct}{\mcitedefaultseppunct}\relax
\EndOfBibitem
\bibitem[Carpenter and Salje(1998)Carpenter, and Salje]{CS98}
Carpenter,~M.~A.; Salje,~E. H.~K. Elastic anomalies in minerals due to
  structural phase transitions. \emph{Eur. J. Mineral.} \textbf{1998},
  \emph{10}, 693--812\relax
\mciteBstWouldAddEndPuncttrue
\mciteSetBstMidEndSepPunct{\mcitedefaultmidpunct}
{\mcitedefaultendpunct}{\mcitedefaultseppunct}\relax
\EndOfBibitem
\bibitem[Li \latin{et~al.}(2013)Li, Zhang, Bithell, Batsanov, Barton, Saines,
  Jain, Howard, Carpenter, and Cheetham]{LZB13}
Li,~W.; Zhang,~Z.; Bithell,~E.~G.; Batsanov,~A.~S.; Barton,~P.~T.;
  Saines,~P.~J.; Jain,~P.; Howard,~C.~J.; Carpenter,~M.~A.; Cheetham,~A.~K.
  Ferroelasticity in a {metal-organic framework perovskite}; towards a new
  class of multiferroics. \emph{Acta Mater.} \textbf{2013}, \emph{61},
  4928\relax
\mciteBstWouldAddEndPuncttrue
\mciteSetBstMidEndSepPunct{\mcitedefaultmidpunct}
{\mcitedefaultendpunct}{\mcitedefaultseppunct}\relax
\EndOfBibitem
\bibitem[Goodenough \latin{et~al.}(1972)Goodenough, Kafalas, and Longo]{GKL72}
Goodenough,~J.~B.; Kafalas,~J.~A.; Longo,~J.~M. In \emph{{Preparative Methods
  in Solid State Chemistry}}; Hagenmuller,~P., Ed.; Academic Press, 1972\relax
\mciteBstWouldAddEndPuncttrue
\mciteSetBstMidEndSepPunct{\mcitedefaultmidpunct}
{\mcitedefaultendpunct}{\mcitedefaultseppunct}\relax
\EndOfBibitem
\bibitem[Leger \latin{et~al.}(1990)Leger, Redon, Andraud, and Pelle]{LRA90}
Leger,~J.~M.; Redon,~A.~M.; Andraud,~C.; Pelle,~F. {Isotropic compression of
  the linear-chain perovskite-type CsCdBr3 up to 20 GPa}. \emph{Phys. Rev. B}
  \textbf{1990}, \emph{41}, 9276\relax
\mciteBstWouldAddEndPuncttrue
\mciteSetBstMidEndSepPunct{\mcitedefaultmidpunct}
{\mcitedefaultendpunct}{\mcitedefaultseppunct}\relax
\EndOfBibitem
\bibitem[Goodenough and Zhou(1998)Goodenough, and Zhou]{GZ98}
Goodenough,~J.~B.; Zhou,~J.~S. Localized to Itinerant Electronic Transitions in
  Transition-Metal Oxides with the Perovskite Structure. \emph{Chem. Mater.}
  \textbf{1998}, \emph{10}, 2980\relax
\mciteBstWouldAddEndPuncttrue
\mciteSetBstMidEndSepPunct{\mcitedefaultmidpunct}
{\mcitedefaultendpunct}{\mcitedefaultseppunct}\relax
\EndOfBibitem
\bibitem[Tian \latin{et~al.}(2019)Tian, Cordes, Quarti, Beljonne, Slawin,
  Zysman-Colman, and Morrison]{TCQ19}
Tian,~J.; Cordes,~D.~B.; Quarti,~C.; Beljonne,~D.; Slawin,~A.~Z.;
  Zysman-Colman,~E.; Morrison,~F.~D. {Stable 6H Organic-Inorganic Hybrid Lead
  Perovskite and Competitive Formation of 6H and 3C Perovskite Structure with
  Mixed A-Cations}. \emph{ACS Appl. Energy Mater.} \textbf{2019}, \emph{2},
  5427\relax
\mciteBstWouldAddEndPuncttrue
\mciteSetBstMidEndSepPunct{\mcitedefaultmidpunct}
{\mcitedefaultendpunct}{\mcitedefaultseppunct}\relax
\EndOfBibitem
\bibitem[Kamminga \latin{et~al.}(2017)Kamminga, de~Wijs, Havenith, Blake, and
  Palstra]{KWH17}
Kamminga,~M.~E.; de~Wijs,~G.~A.; Havenith,~R.~A.; Blake,~G.~R.; Palstra,~T.~M.
  {The Role of Connectivity on Electronic Properties of Lead Iodide
  Perovskite-Derived Compounds}. \emph{Inorg. Chem.} \textbf{2017}, \emph{56},
  8408\relax
\mciteBstWouldAddEndPuncttrue
\mciteSetBstMidEndSepPunct{\mcitedefaultmidpunct}
{\mcitedefaultendpunct}{\mcitedefaultseppunct}\relax
\EndOfBibitem
\bibitem[Trots and Myagkota(2008)Trots, and Myagkota]{TM08}
Trots,~D.~M.; Myagkota,~S.~V. {High-temperature structural evolution of caesium
  and rubidium triiodoplumbates}. \emph{J. Phys. Chem. Sol.} \textbf{2008},
  \emph{69}, 2520\relax
\mciteBstWouldAddEndPuncttrue
\mciteSetBstMidEndSepPunct{\mcitedefaultmidpunct}
{\mcitedefaultendpunct}{\mcitedefaultseppunct}\relax
\EndOfBibitem
\bibitem[Chung \latin{et~al.}(2012)Chung, Song, Im, Androulakis, Malliakas, Li,
  Freeman, Kenney, and Kanatzidis]{CSI12}
Chung,~I.; Song,~J.~H.; Im,~J.; Androulakis,~J.; Malliakas,~C.~D.; Li,~H.;
  Freeman,~A.~J.; Kenney,~J.~T.; Kanatzidis,~M.~G. {CsSnI3: Semiconductor or
  Metal? High Electrical Conductivity and Strong Near-Infrared
  Photoluminescence from a Single Material. High Hole Mobility and
  Phase-Transitions}. \emph{J. Am. Chem. Soc.} \textbf{2012}, \emph{134},
  8579\relax
\mciteBstWouldAddEndPuncttrue
\mciteSetBstMidEndSepPunct{\mcitedefaultmidpunct}
{\mcitedefaultendpunct}{\mcitedefaultseppunct}\relax
\EndOfBibitem
\bibitem[Stoumpos \latin{et~al.}(2017)Stoumpos, Mao, Malliakas, and
  Kanatzidis]{SMM17b}
Stoumpos,~C.~C.; Mao,~L.; Malliakas,~C.~D.; Kanatzidis,~M.~G. {Structure-Band
  Gap Relationships in Hexagonal Polytypes and Low-Dimensional Structures of
  Hybrid Tin Iodide Perovskites}. \emph{Inorg. Chem.} \textbf{2017}, \emph{56},
  56\relax
\mciteBstWouldAddEndPuncttrue
\mciteSetBstMidEndSepPunct{\mcitedefaultmidpunct}
{\mcitedefaultendpunct}{\mcitedefaultseppunct}\relax
\EndOfBibitem
\bibitem[Guo \latin{et~al.}(2019)Guo, Yang, and Lightfoot]{GYL19}
Guo,~Y.~Y.; Yang,~L.~J.; Lightfoot,~P. {Three New Lead Iodide Chain Compounds,
  APbI$_3$, Templated by Molecular Cations}. \emph{Crystals} \textbf{2019},
  \emph{9}\relax
\mciteBstWouldAddEndPuncttrue
\mciteSetBstMidEndSepPunct{\mcitedefaultmidpunct}
{\mcitedefaultendpunct}{\mcitedefaultseppunct}\relax
\EndOfBibitem
\bibitem[S{\o}nden{\aa} \latin{et~al.}(2007)S{\o}nden{\aa}, St{\o}len,
  Ravindran, Grande, and Allan]{SSR07}
S{\o}nden{\aa},~R.; St{\o}len,~S.; Ravindran,~P.; Grande,~T.; Allan,~N.~L.
  {Corner- versus face-sharing octahedra in AMnO$_3$ perovskites (A = Ca, Sr,
  and Ba)}. \emph{Phys. Rev. B} \textbf{2007}, \emph{75}, 184105\relax
\mciteBstWouldAddEndPuncttrue
\mciteSetBstMidEndSepPunct{\mcitedefaultmidpunct}
{\mcitedefaultendpunct}{\mcitedefaultseppunct}\relax
\EndOfBibitem
\bibitem[Daub \latin{et~al.}(2018)Daub, Ketterer, and Hillebrecht]{DKH18}
Daub,~M.; Ketterer,~I.; Hillebrecht,~H. {Syntheses, Crystal Structures, and
  Optical Properties of the Hexagonal Perovskites Variants ABX3 (B = Ni, A =
  Gu, FA, MA, X = Cl, Br; B = Mn, A = MA, X = Br)}. \emph{Z. Anorg. Allg.
  Chem.} \textbf{2018}, \emph{644}, 280\relax
\mciteBstWouldAddEndPuncttrue
\mciteSetBstMidEndSepPunct{\mcitedefaultmidpunct}
{\mcitedefaultendpunct}{\mcitedefaultseppunct}\relax
\EndOfBibitem
\bibitem[Cordero \latin{et~al.}(2017)Cordero, Craciun, and Trequattrini]{CCT17}
Cordero,~F.; Craciun,~F.; Trequattrini,~F. {Ionic Mobility and Phase
  Transitions in Perovskite Oxides for Energy Applications}. \emph{Challenges}
  \textbf{2017}, \emph{8}, 5\relax
\mciteBstWouldAddEndPuncttrue
\mciteSetBstMidEndSepPunct{\mcitedefaultmidpunct}
{\mcitedefaultendpunct}{\mcitedefaultseppunct}\relax
\EndOfBibitem
\bibitem[Cordero \latin{et~al.}(2018)Cordero, Craciun, Trequattrini,
  Imperatori, Paoletti, and Pennesi]{CCT18}
Cordero,~F.; Craciun,~F.; Trequattrini,~F.; Imperatori,~P.; Paoletti,~A.~M.;
  Pennesi,~G. {Competition between Polar and Antiferrodistortive Modes and
  Correlated Dynamics of the Methylammonium Molecules in MAPbI$_3$}. \emph{J.
  Phys. Chem. Lett.} \textbf{2018}, \emph{9}, 4401\relax
\mciteBstWouldAddEndPuncttrue
\mciteSetBstMidEndSepPunct{\mcitedefaultmidpunct}
{\mcitedefaultendpunct}{\mcitedefaultseppunct}\relax
\EndOfBibitem
\bibitem[Fabini \latin{et~al.}(2016)Fabini, Stoumpos, Laurita, Kaltzoglou,
  Kontos, Falaras, Kanatzidis, and Seshadri]{FSL16}
Fabini,~D.~H.; Stoumpos,~C.~C.; Laurita,~G.; Kaltzoglou,~A.; Kontos,~A.~G.;
  Falaras,~P.; Kanatzidis,~M.~G.; Seshadri,~R. {Reentrant Structural and
  Optical Properties and Large Positive Thermal Expansion in Perovskite
  Formamidinium Lead Iodide}. \emph{Angew. Chem. Int. Ed.} \textbf{2016},
  \emph{55}, 15392\relax
\mciteBstWouldAddEndPuncttrue
\mciteSetBstMidEndSepPunct{\mcitedefaultmidpunct}
{\mcitedefaultendpunct}{\mcitedefaultseppunct}\relax
\EndOfBibitem
\bibitem[Wu and Lin(2005)Wu, and Lin]{WL05}
Wu,~C.~D.; Lin,~W. {Highly Porous, Homochiral Metal-Organic Frameworks:
  Solvent-Exchange-Induced Single-Crystal to Single-Crystal Transformations}.
  \emph{Angew. Chem. Int. Ed.} \textbf{2005}, \emph{44}, 1958\relax
\mciteBstWouldAddEndPuncttrue
\mciteSetBstMidEndSepPunct{\mcitedefaultmidpunct}
{\mcitedefaultendpunct}{\mcitedefaultseppunct}\relax
\EndOfBibitem
\bibitem[Dastidar \latin{et~al.}(2017)Dastidar, Hawley, Dillon,
  Gutierrez-Perez, Spanier, and Fafarman]{DHD17}
Dastidar,~S.; Hawley,~C.~J.; Dillon,~A.~D.; Gutierrez-Perez,~A.;
  Spanier,~J.~E.; Fafarman,~A.~T. {Quantitative Phase-Change Thermodynamics and
  Metastability of Perovskite-Phase Cesium Lead Iodide}. \emph{J. Phys. Chem.
  Lett.} \textbf{2017}, \emph{8}, 1278\relax
\mciteBstWouldAddEndPuncttrue
\mciteSetBstMidEndSepPunct{\mcitedefaultmidpunct}
{\mcitedefaultendpunct}{\mcitedefaultseppunct}\relax
\EndOfBibitem
\bibitem[Yadavalli \latin{et~al.}(2020)Yadavalli, Dai, Hu, Dong, Li, Zhou, Zia,
  and Padture]{YDH20}
Yadavalli,~S.~K.; Dai,~Z.; Hu,~M.; Dong,~Q.; Li,~W.; Zhou,~Y.; Zia,~R.;
  Padture,~N.~P. {Mechanisms of Exceptional Grain Growth and Stability in
  Formamidinium Lead Triiodide Thin Films for Perovskite Solar Cells}.
  \emph{Acta Mater.} \textbf{2020}, \relax
\mciteBstWouldAddEndPunctfalse
\mciteSetBstMidEndSepPunct{\mcitedefaultmidpunct}
{}{\mcitedefaultseppunct}\relax
\EndOfBibitem
\bibitem[Arakcheeva \latin{et~al.}(2017)Arakcheeva, Svitlyk, Koll{\'a}r,
  N{\'a}fr{\'a}di, Forr{\'o}, and Horv{\'a}th]{ASK17}
Arakcheeva,~A.; Svitlyk,~V.; Koll{\'a}r,~M.; N{\'a}fr{\'a}di,~B.;
  Forr{\'o},~L.; Horv{\'a}th,~E. {High-pressure transformation of MAPbI$_3$ :
  role of the noble-gas medium}. \emph{Acta Cryst. A} \textbf{2017}, \emph{70},
  C1416\relax
\mciteBstWouldAddEndPuncttrue
\mciteSetBstMidEndSepPunct{\mcitedefaultmidpunct}
{\mcitedefaultendpunct}{\mcitedefaultseppunct}\relax
\EndOfBibitem
\bibitem[You \latin{et~al.}(2014)You, Yang, Hong, Song, Meng, Liu, Jiang, Zhou,
  Chang, Li, and Yang]{YYH14}
You,~J.; Yang,~Y.~M.; Hong,~Z.; Song,~T.~B.; Meng,~L.; Liu,~Y.; Jiang,~C.;
  Zhou,~H.; Chang,~W.~H.; Li,~G. \latin{et~al.}  {Moisture assisted perovskite
  film growth for high performance solar cells}. \emph{Appl. Phys. Lett.}
  \textbf{2014}, \emph{105}, 183902\relax
\mciteBstWouldAddEndPuncttrue
\mciteSetBstMidEndSepPunct{\mcitedefaultmidpunct}
{\mcitedefaultendpunct}{\mcitedefaultseppunct}\relax
\EndOfBibitem
\bibitem[Singh and Miyasaka(2018)Singh, and Miyasaka]{SM18}
Singh,~T.; Miyasaka,~T. {Stabilizing the Efficiency Beyond 20
  Cation Perovskite Solar Cell Fabricated in Ambient Air under Controlled
  Humidity}. \emph{Adv. Energy Mater.} \textbf{2018}, \emph{8}, 1700677\relax
\mciteBstWouldAddEndPuncttrue
\mciteSetBstMidEndSepPunct{\mcitedefaultmidpunct}
{\mcitedefaultendpunct}{\mcitedefaultseppunct}\relax
\EndOfBibitem
\bibitem[Anderson and Morgan(1964)Anderson, and Morgan]{AM64}
Anderson,~P.~J.; Morgan,~P.~L. {Effects of water vapour on sintering of MgO}.
  \emph{Trans. Faraday Soc.} \textbf{1964}, \emph{60}, 930\relax
\mciteBstWouldAddEndPuncttrue
\mciteSetBstMidEndSepPunct{\mcitedefaultmidpunct}
{\mcitedefaultendpunct}{\mcitedefaultseppunct}\relax
\EndOfBibitem
\bibitem[Charles \latin{et~al.}(2017)Charles, Dillon, Weber, Islama, and
  Weller]{CDW17}
Charles,~B.; Dillon,~J.; Weber,~O.~J.; Islama,~M.~S.; Weller,~M.~T.
  {Understanding the stability of mixed A-cation lead iodide perovskites}.
  \emph{J. Mater. Chem A} \textbf{2017}, \emph{5}, 22495\relax
\mciteBstWouldAddEndPuncttrue
\mciteSetBstMidEndSepPunct{\mcitedefaultmidpunct}
{\mcitedefaultendpunct}{\mcitedefaultseppunct}\relax
\EndOfBibitem
\bibitem[Weller \latin{et~al.}(2015)Weller, Weber, Frost, and Walsh]{WWF15}
Weller,~M.~T.; Weber,~O.~J.; Frost,~J.~M.; Walsh,~A. {Cubic Perovskite
  Structure of Black Formamidinium Lead Iodide, $\alpha
  -$[HC(NH$_2$)$_2$]PbI$_3$, at 298~K}. \emph{J. Phys. Chem. Lett.}
  \textbf{2015}, \emph{6}, 3209\relax
\mciteBstWouldAddEndPuncttrue
\mciteSetBstMidEndSepPunct{\mcitedefaultmidpunct}
{\mcitedefaultendpunct}{\mcitedefaultseppunct}\relax
\EndOfBibitem
\bibitem[Weber \latin{et~al.}(2018)Weber, Ghosh, Gaines, Henry, Walker, Islam,
  and Weller]{WGG18}
Weber,~O.~J.; Ghosh,~D.; Gaines,~S.; Henry,~P.~F.; Walker,~A.~B.; Islam,~M.~S.;
  Weller,~M.~T. {Phase Behavior and Polymorphism of Formamidinium Lead Iodide}.
  \emph{Chem. Mater.} \textbf{2018}, \emph{30}, 3768\relax
\mciteBstWouldAddEndPuncttrue
\mciteSetBstMidEndSepPunct{\mcitedefaultmidpunct}
{\mcitedefaultendpunct}{\mcitedefaultseppunct}\relax
\EndOfBibitem
\bibitem[Doherty \latin{et~al.}(2020)Doherty, Winchester, Macpherson,
  Johnstone, Pareek, Tennyson, Kosar, Kosasih, Anaya, J., G., Wong, Mad{\'e}o,
  Chiang, Park, Jung, Petoukhoff, Divitini, Man, Ducati, Walsh, Midgley, Dani,
  and Stranks]{DWM20}
Doherty,~T.~S.; Winchester,~A.~J.; Macpherson,~S.; Johnstone,~D.~N.;
  Pareek,~V.; Tennyson,~E.~M.; Kosar,~S.; Kosasih,~F.~U.; Anaya,~M.; J.,~M.~A.
  \latin{et~al.}  {Performance-limiting nanoscale trap clusters at grain
  junctions in halide perovskites}. \emph{Nature} \textbf{2020}, \emph{580},
  360\relax
\mciteBstWouldAddEndPuncttrue
\mciteSetBstMidEndSepPunct{\mcitedefaultmidpunct}
{\mcitedefaultendpunct}{\mcitedefaultseppunct}\relax
\EndOfBibitem
\end{mcitethebibliography}
\newpage

\section{Supplementary Information}

\newpage

\section{Synthesis}

Formamidine acetate (99\%), HI (57 wt\% in water), $\gamma $%
-butyrolactone (GBL 99\%), PbI$_{2}$ (99\%), and other solvents were purchased
from Sigma-Aldrich and were used as received without further purification.

We first describe the synthesis of the powders for samples FAPI \#4 and 5.
The preliminary synthesis of CH(NH$_{2}$)$_{2}$I (FAI) was accomplished
by reacting 50~mL hydroiodic acid (0.38~mol) and 25~g formamidine acetate
(0.24~mol) in a 250~mL round bottomed flask at 0~${}^{\circ }$C for 6~h under
N$_{2}$ atmosphere with stirring. The precipitates were recovered by reduced
pressure evaporating the solutions at 65~${}^{\circ }$C for about 3~h. The
product was washed with absolute ethyl alcohol and diethyl ether until the
washings were colorless and finally dried at 60~${}^{\circ }$C in a vacuum
oven for 24~h. (E.A. FAI calculated: C 6.98, H 2.93, N
16.29, found: C 6.98, H 2.96, N 16.05).

Synthesis of FAPI $\delta $ phase: equimolar mixture (1.2~M) of prepared FAI
and PbI$_{2}$ were dissolved under vigorous stirring in GBL at $\sim 50$~$%
{}^{\circ }$C until a clear solution was obtained; then a small amount of
toluene was added to the solution at room temperature, the yellow precipitates
were recovered by centrifugation, washed with toluene and dried under vacuum.

For the synthesis of the powder for sample FAPI \#7 a method in water was
followed:\cite{JBN19} 0.32~g of formamidine acetate (Aldrich) (3~mmol) were
added to 2~mL of HI (aq solution 57\% ACROS) (9~mmol), then 1.14~g of Pb(CH$%
_{3}$CO$_{2}$)$_{2}\cdot $3H$_{2}$O (CarloErbaReagents RP, 3~mmol) were
added under stirring; the yellow suspension was put in a ultrasonic bath at
room temperature for 10~min, then was filtered and washed with ethyl acetate
and dried under vacuum first for 4~h at 50~${}^{\circ }$C and then at room
temperature overnight, obtaining pure $\delta -$FAPI according to the XRD
analysis.

\section{Pressing}

\begin{figure}[ht]
\includegraphics[width=9 cm]{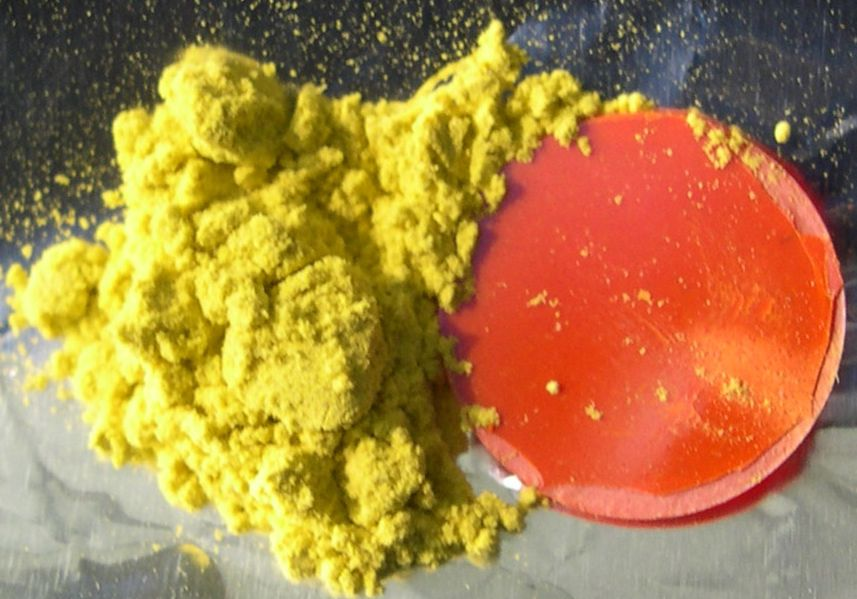}
\caption{Powder of $\protect\delta -$FAPI before and after pressing into a
disc at 0.6~GPa for a few minutes.}
\label{S1}
\end{figure}

The yellow powders of $\delta -$FAPI were pressed at 5~tons in discs of
13~mm of diameter (0.62~GPa)\ and at 10~tons in bars$\ $with flat surface $%
40\times 6$~mm$^{2}$ (0.2~GPa). The maximum pressure was reached gradually
in about 2~min and kept for about 4~min. Figure \ref{S1} shows the pristine
yellow powder of $\delta -$FAPI and a disc pressed from the same powder.

\begin{figure}[ht]
\includegraphics[width=9 cm]{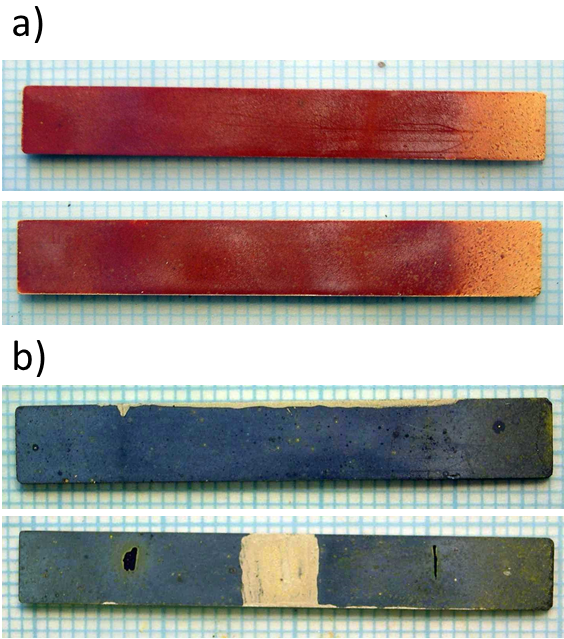}
\caption{Bar FAPI~\#7 just after being pressed from $\protect\delta -$FAPI
powder and after a series of anelastic measurements during which it
transformed into $\protect\alpha $ phase. An edge and the center of a face
were made conductive with Ag paint. The black signs were made with a
permanent marker at the nodal lines of the first flexural mode.}
\label{S2}
\end{figure}

Figure \ref{S2} (a) shows the bar FAPI \#7 just after being pressed.

The non uniform color reflects the non uniform distribution of the powder in
the die, with the lighter extreme corresponding to smaller initial
thickness and hence density and local pressure. After the initial heating in
7~mbar of H$_{2}$O the sample transformed into the black perovskite $\alpha $
phase, but the non uniform yellow hue and the XRD spectrum indicate that
some PbI$_{2}$ had formed.

\section{X-ray diffraction}

X-ray diffraction measurements (XRD) were performed on a Panalytical
Empyrean Diffractometer, using the K$_{\alpha }$ fluorescence line of a
Cu-anode emitting tube as X-ray source. Bragg Brentano configuration was
used as incident optical pathway (divergent slits) and a solid-state hybrid
Pix'cel 3D detector, working in 1D linear mode, accomplished the detection.

None of the samples for the dielectric and anelastic experiments was analyzed
with X--rays just after pressing, in order to reduce as much as possible any
decomposition from the humidity that rapidly coalesces at grain boundaries.\cite{CCT19}
Therefore, in order to search for XRD evidence of the orange 4H and 6H phases, we pressed
other discs (FAPI \#9) in conditions identical to those for the dielectric experiments,
as shown in Fig. \ref{S1}.

\begin{figure*}[h]
\includegraphics[width=17 cm]{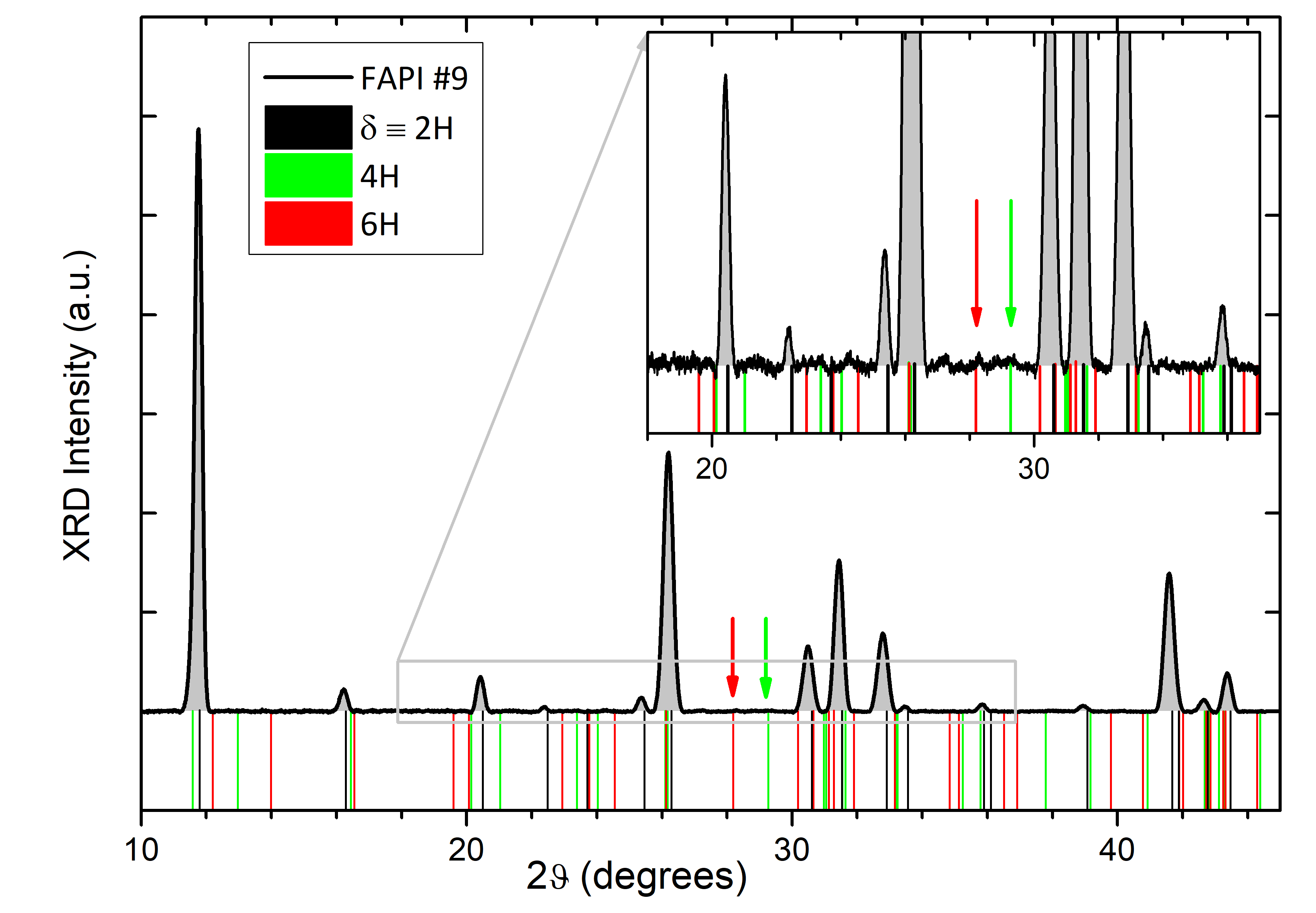}
\caption{XRD pattern of FAPI \#9 with the expected positions of the reflections
from the 2H, 4H and 6H phases.}
\label{fig-XRD-FAPI9}
\end{figure*}

XRD measurements were performed with fast data scans (0.5~s/point,
overall measurement time 1.5~h.) in the $5^\circ - 45^\circ$ angular range except
for the slower scan of Fig. \ref{fig-XRD-FAPI9} (1~s/point with step in $2 \theta$
of $0.002^\circ$, overall measurement time 5~h 40~min).

In the absence of literature on the 4H and 6H phases of pure FAPI, we compare
our spectra with those simulated with VESTA
[Momma,~K.; Izumi,~F. "VESTA 3 for three-dimensional visualization of crystal,
  volumetric and morphology data" \emph{J. Appl. Cryst.}\emph{44}, 1272 (2011)]
using the Crystallographic
Information Files (CIF) of FA$_{0.85}$MA$_{0.15}$PbI$_{3}$ from Refs.
\citenum{GZS17,Gra18}. Even though it is expected that the partial substitution of
FA with MA reduces the cell volume, the spectra of the 2H phase simulated in
this manner are practically identical with those of the $\delta \equiv $2H,
suggesting that also the reflections of the 4H and 6H phases of the MA--substituted
material are representative of those of pure FAPI.
Figure S\ref{fig-XRD-FAPI9} shows the XRD pattern of FAPI\#9.
As expected, the $\delta \equiv $2H phase is clearly detected (black column bars
also reported in Table 1). Most of the expected 4H and 6H reflections with intensities
$>1\%$ are close to each other and to those of the dominant 2H phase, but
the inset highlights the $18^\circ <$ $2\theta <$
$37^\circ$ region where it is possible to distinguish two peaks that match
two isolated reflections of the 4H and 6H phases (red and green arrows).

It is possible that the difficulty of distinguishing clear reflections from additional
phases in orange samples arises from their lower crystallinity with respect to the
$\delta $ phase, as discussed for the $\alpha $ phase in the main text.

\begin{table*}
\caption{Main reflections of $\delta -$FAPI simulated with VESTA
using the CIF from Ref. \citenum{SMK13c} and corresponding to the black bars
in Fig. \ref{fig-XRD-FAPI9}.}
\begin{tabular}{ccc}
$2\theta $ & Miller indexes $(hkl)$ & Relative Intensities (\%) \\
11.790 & 100 & 100 \\
16.281 & 101 / 10-1 & 7.564 / 6.665 \\
20.494 & 2-10 & 3.544\\
22.484 & 002 / 00-2 & 2.381 / 2.663\\
25.460 & 102 & 5.898\\
26.284 & 201 / 20-1 & 28.890 / 29.340\\
30.606 & 2-12 / 2-1-2 & 17.100 / 16.620\\
31.535 & 3-10 & 20.529\\
32.903 & 202 / 20-2 & 18.731 / 18.803\\
33.559 & 3-11 / 3-1-1 & 2.063 / 2.009\\
35.892 & 300 & 1.114\\
36.112 & 103 / 10-3 & 1.149 / 1.057\\
39.077 & 3-1-2 / 3-12 & 2.570 / 2.491\\
41.683 & 4-20 & 18.138\\
41.877 & 203 / 20-3 & 4.947 / 5.052\\
42.768 & 302 / 30-2 & 4.530 / 4.466\\
43.470 & 4-10 & 6.943\\
\end{tabular}
\end{table*}

\end{document}